\documentclass[10pt,twocolumn]{article}
\usepackage{amsmath}
\usepackage{amssymb}
\usepackage{epsfig}

\begin{document}

\setlength{\baselineskip}{12pt}

\title{An Auction-driven Self-organizing Cloud Delivery Model}
\author{Dan C. Marinescu and Ashkan Paya\\
Computer Science Division \\
Department of Electrical Engineering and Computer Science \\
University of Central Florida, Orlando, FL 32816, USA \\
Email:dcm@cs.ucf.edu, ashkan\_paya@knights.ucf.edu \\ \\
John P. Morrison and Philip Healy \\
Computer Science Department \\
University College Cork. Cork, Ireland \\
Email: (j.morrison, p.healy)@cs.ucc.ie}

\maketitle

\begin{abstract}
The three traditional cloud delivery models -- IaaS, PaaS, and SaaS -- constrain access to cloud resources by hiding their raw functionality and forcing us to use them indirectly via a restricted set of actions.  Can we introduce a new delivery model, and, at the same time, support improved security, a higher degree of assurance, find relatively simple solutions to the hard cloud resource management problems, eliminate some of the inefficiencies related to resource virtualization, allow the assembly of clouds of clouds, and, last but not least, minimize the number of interoperability standards?

We sketch a self-organizing architecture for very large compute clouds composed of many-core processors and heterogeneous coprocessors. We discuss how self-organization will address each of the challenges described above. The approach is {\em bid-centric}. The system of heterogeneous cloud resources is dynamically, and autonomically, configured to bid to meet the needs identified in a high-level task or service specification. When the task is completed, or the service is retired, the resources are released for subsequent reuse.

Our approach mimics the process followed by individual researchers who, in response to a call for proposals released by a funding agency, organize themselves in groups of various sizes and specialities. If the bid is successful, then the group carries out the proposed work and releases the results.  After the work is completed, individual researchers in the group disperse, possibly joining other groups or submitting individual bids in response to other proposals. Similar protocols are common to other human activities such as procurement management.
\end{abstract}

\section{Motivation}
\label{Motivation}
\medskip
From the beginning, CSPs (Cloud Service Providers) made their cloud offerings available all based on the three delivery models,  SaaS (Software as a Service), PaaS (server as a Service), and IaaS (Infrastructure as a Service),  that represented the then state of the art.  As cloud computing gained traction, cloud vendors adapted their products and services to fit into the three delivery models available to them.

As a result, as the models became more used, they became the dominant design patterns and, more subtly, they became the dominant way of thinking about cloud computing. This has given rise to a constrained view of the possibilities afforded by the cloud. In this paper, we attempt to break with the traditional mode of thinking and to look afresh at how cloud services are delivered.

Today's cloud computing landscape is partitioned into Saas, PaaS, IaaS. These models provide very different types of services and have different capabilities and internal structure. The emergence of the three models during the first decade of the new millennium was well-justified, as they respectively targeted different types of application and distinct user groups. These delivery models were initially motivated as sales channels  with which CSPs such as  Google, Microsoft, and Amazon could target integrators, Independent Software Vendors (ISVs) and consumers. This gave rise to a strong ecosystem in which these models were the only alternatives.

With SaaS, software becomes an appliance, the administration of which is outsourced. Typically, little or no technical expertise is required to consume a SaaS offering. This makes it easy to sell, the end-user simply needs to trust the software vendor and the cloud provider to deliver the service competently. PaaS places some responsibility on the vendor/consumer to manage the service lifecycle, but without requiring them to engage in low-level systems adminstration and provisioning. IaaS provides the consumer with resources from which he can create familiar environments. These can be provisioned and customized to meet specific requirements. A high-level of customization can yield competitive advantage without associated costs of ownership, however, IaaS demands a high degree of administrative competence and places the onus on the consumer to shoulder the ensuant responsibility.   Restricting access to services supported by software developed in-house, as in the case of SaaS, or to software installed after the verification and approval of the CSP, as in the case of PaaS, limits the security exposure of the CSP. The internal resource management policies and mechanisms to implement these policies are simpler for both SaaS and PaaS than the ones for IaaS.

Typically, a cloud stack is constructed in layers; each layer being composed of building blocks from the layer on which it sits.  At the bottom of the stack is the infrastructure layer that employs virtualization to deliver physical resources to consumers. The virtualization process, by its nature, leads to resource fragmentation and hence to inefficiencies. Consider two virtual machines running on the same physical host, where one is overloaded and the other is underloaded. The boundary defined by the virtualization process prohibits the overloaded machine from exploiting the physical resources allocated to the unloaded machine and hypervisor. Since the clustering of virtual machines is the most common method of achieving horizontal scalability, this inefficiency propagates to the higher layers in the stack. Moreover, new inefficiencies arise in the upper layers of the stack in the form of contention for shared resources in a multi-tenancy environment. Current approached for dealing with shared resource contention, such as VM migration to achieve an optimal heterogeneous mix, are purely reactive. Could a more proactive approach be taken to shared resource management?

The Openstack's baremetal driver attempts to address the inefficiencies caused by virtualization fragmentation and hence could be argued to represent an emerging fourth cloud delivery model - MaaS (Metal-as-a-Service) \cite{MAAS}.  However, even more so than IaaS, MaaS requires a high degree of technical skill to deploy and manage and this makes it, at least at present, difficult to widely consume. Furthermore, unless a physical resource is utilized to its fullest, there is still a resource utilization issue. However, it is now passed from the CSP to the consumer. Is it possible to deliver these physical resource without precipitating these inefficiencies?

The proliferation of CSPs has given rise to many different APIs for performing similar tasks, such as IaaS provisioning. Different providers innovate in different ways, however, these differences are not conducive to implementing portable or interoperable applications. Would a more declarative means of specifying services free us from the mechanics of how they are provisioned and hence foster portability and interoperability?

The "walled Gardens" of the current CSPs are analogous to the fragmented state of computer networks before the advent of the Internet. Since the introduction of the Internet Protocol, hardware and software has undergone a dramatic evolution allowing the Internet to became a critical infrastructural component of society. Today the Internet supports the Web, electronic commerce, data streaming,  and countless other applications including cloud computing.

With the glue provided by the Internet Protocol, the Internet was free to develop organically; unconstrained by top-down regulation. Today, clouds are islands among which communication is difficult. Could a new delivery model bridge these islands and accelerate the development of the cloud ecosystem?



The work reported in this paper attempts to answer some the questions posed above. It is motivated by the desire to improve and enrich the cloud computing landscape and by the desire to identify a disruptive technology that addresses some of the very hard problems at the core of today's cloud computing infrastructure and service delivery models. The architecture we propose has its own limitations, it cannot eliminate all the inefficiencies inherent to virtualization, requires the development of new families of algorithms for resource management and the development of new software. On balance this approach has compelling advantages as we shall see in Section \ref{Architecture}.

\section{Self-organization - a Disruptive Technology for Cloud Computing}
\label{SelfManagement}
\medskip
Some of the challenges to the cloud delivery models discussed in Section \ref{Motivation} are amply documented in the literature. Security and privacy \cite{Auty10,Balduzzi12,CSA11,Garfinkel05,Gruschka10,Li10,Pearson10,Price08}, the virtualization overhead \cite{Chandra03,Marinescu13,Menon05}, and sustainability \cite{Chang10,Paya13,Van10} are major concerns motivating research for new architectural solutions for cloud computing \cite{Celesti10,Sarathy10}.

A review of some of the challenges faced by cloud computing hints that computer  clouds are complex systems \cite{Abbott07,Sommerville12}. A complex system is one with a very large number components, each with distinct characteristics, and many interaction channels among individual components. Four groups of actors are involved in cloud computing: (i) the CSP infrastructure consisting of possibly millions of compute and storage servers and an interconnection network and the stack of software running on each of these systems; (ii) a very large population of individual and corporate users with different level of expertise, expectations, applications, and constraints; (iii) the regulators, the government agencies that enforce the rules governing the business; (iv) and, last but not least, the physical environment, including the networks supporting user access and the power grid supplying the energy for powering the systems, the heating and the cooling. These components interact with one another often in unexpected ways; a faulty error recovery procedure triggered by the power failure of a few systems could cause a chain reaction and shut down an entire data center, thus affecting a very large user population.

Today's clouds are designed and engineered using technics suitable for small-scale deterministic systems rather than complex systems with non-deterministic behavior. The disruptive technology we advocate for a cloud infrastructure is based on {\it self-organization} and {\it self-management}.

Informally, self-organization means synergetic activities of elements when no single element acts as a coordinator and the global patterns of behavior are distributed \cite{Gell-Mann88,Schuster07}. The intuitive meaning of self-organization is captured by the observation of Alan Turing \cite{Turing52}: ``global order can arise from local interactions.'' More recent concepts such as {\it autonomic computing} introduced by IBM and {\it organic computing}  have some intersection with self-organization and imply autonomy of individual system components and intelligent behavior.

Self-organization is prevalent in nature; for example, this process is responsible for molecular self-assembly, for self-assembly of monolayers, for the formation of liquid and colloidal crystals, and in many other instances. Spontaneous folding of proteins and other biomacromolecules, the formation of lipid bilayer membranes, the flocking behavior of different species, the creation of structures by social animals, are all manifestation of self-organization of biological systems. Inspired by biological systems, self-organization was proposed for the organization of different types of computing and communication systems \cite{Hopfield82, Marinescu10}, including sensor networks, for space exploration \cite{Hinchey06}, and even for economical systems \cite{Krugman96}.
A number of studies of self-organization in physical systems and the mechanisms to control such systems have been published recently \cite{Lu11, Noel13}.

Though  the virtues of self-management have long been recognized \cite{Ardagna07, Ardagna11, Gupta09}, there is, to our knowledge, no cloud computing infrastructure, or large-scale computing or communication system based on self-organizing principles.
Some of the mechanisms used in our model have been discussed in the literature e.g., \cite{Azambuja10, Bernstein11,Brandic10,Dutreilh10,Gmach09,Lim09}, others have been incorporated in different cloud architectures \cite{Arrasjid13,Sugerman01}.

\section{A Cloud Architecture Based on Auctions and Self-management}
\label{Architecture}

The  model we propose uses a market approach based on combinatorial auctions.  {\it Combinatorial auctions} \cite{Cramton06, Stokely10} allow participants to bid on bundles of items or {\it packages} e.g., combinations of CPU cycles, main memory, secondary storage, I/O and network bandwidth. The auctions provide a relatively simple, scalable, and tractable solution to cloud resource allocation, eliminate the need for admission control policies, which require some information about the global state of the system and, most importantly, allow the service to be tailored to the specific privacy, security, and Quality of Service (QoS) needs of each application. At this time, Amazon Web Services (AWS) support the so called {\it spot instances}, based on a market-oriented pricing strategy.

The  application of market-oriented mechanisms \cite{Marinescu08,Marinescu09,Stokely10} and their advantages over the other basic mechanisms implementing resource management policies in large-scale systems have been analyzed in the literature. Control theory \cite{Kalyvianaki09,Kusic08}, machine learning \cite{Kephart07,Tung07}, and utility-based methods require a detailed model of the system, are not scalable, and typically support.  If no bid exists for a service request then the request cannot be accepted. This procedure acts as an effective admission control. When a bid is generated, the resources are guaranteed to be available. Delivering on that bid is based solely on bidder's local state, so the ability to quantify and to satisfy QoS constraints can be established with a higher degree of assurance. Energy optimization decisions such as:  how to locate servers using only green energy  \cite{Lin12}, how to ensure that individual servers operate within the boundaries of optimal energy consumption \cite{Gandhi11,Gandhi12,Gandhi12a,Paya13}, when and how to switch lightly loaded servers to a sleep state \cite{Gandhi12b} will also be based on local information in an auction-based system. Thus, basing decisions on local information will be accurate and will not require a sophisticated system model nor a large set of parameters that cannot be easily obtained in a practical environment.

The basic tenets of the model we propose are: autonomy of individual components, self-awareness, and intelligent
behavior of individual components. Self-awareness enables servers to create coalitions of peers working in
concert to respond to the needs of an application, rather that offering a menu of a limited number of resource
packages, as is the case with AWS. Again, {\it global order results from local interactions}.

\begin{figure}[!ht]
\begin{center}
\includegraphics[width=8.5cm]{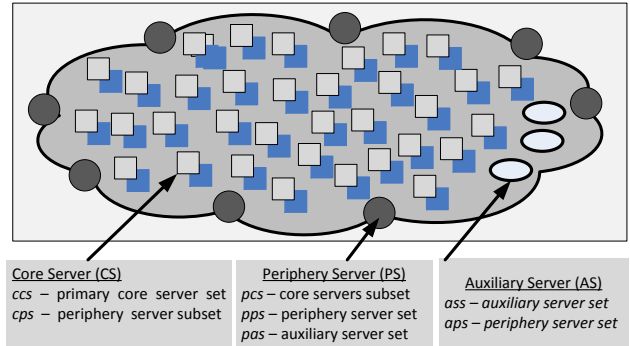}
\end{center}
\caption{The cloud consists of a core and a periphery populated by core and periphery servers, respectively; a
group of auxiliary servers support internal services. Some of the data structures used for self-awareness are shown: (i)  {\it ccl} - the set of primary contacts for a core server and {\it cps} - the subset of periphery servers known to core server; (ii) {\it pcs} - the subset of core servers known to the periphery server, {\it pps} - the set of all periphery servers, and {\it pas} - the set of all auxiliary servers; and (iii)  {\it aas} - the set of auxiliary servers and {\it aps} -the set of periphery servers known to an auxiliary server. These data structures are populated during the initial system configuration and updated throughout its lifetime.}
\label{SOC}
\end{figure}

{\bf Cloud organization.} We distinguish between a core server and a cloud periphery server, Figure \ref{SOC}. The {\it core} consists of a very large number of computational and storage servers, CS, dedicated to the cloud mission - the provision of services to the user community. The core servers are typically heterogeneous; co-processors and/or GPUs are attached to multi/many core processors. A relatively small number of {\it periphery servers}, PS, known to the outside world, act as access points to the core and as the nerve center of the system. There are also a few,  auxiliary servers, AS, providing internal cloud services.

\smallskip

{\bf Operation modes.} The cloud computing infrastructure is reconfigurable; a core server is an autonomous entity that can operate in several modes:

\smallskip

\noindent (i) M1: the server is configured to support multiple virtual machines, as in existing clouds.

\smallskip

\noindent (ii) M2: the server is either the leader or a member of coalition of core servers running under the control of one OS designed to support Single System Image (SSI) operation. This mode will be able to support data-intensive applications which require a very large number of concurrent threads/processes of one application.

\smallskip

\noindent (iii) M3: the server is either the leader or a member of coalition of core servers where each server runs under the control of one of the operating systems supported by the cloud. This mode is of particular interest for data-intensive applications with a complex workflow involving multiple applications and under strict privacy and security constraints; it is inspired by the clusters supported by AWS and by the MaaS (Metal as a Service) model \cite{MAAS}.

\smallskip
\noindent (iv) M4: operation as a storage server.

\smallskip

{\bf Cloud core; the self-configuring monitor (ScM).} The cloud core consists of computational and storage
servers providing services to the user community. To address the natural tension between self-awareness and the desire to maintain minimum state information, each core server has a relatively small set of {\it primary contacts},  and, when necessary, it has access to a much larger pool of {\it secondary contacts.}  {\it Secondary contacts}, are core servers that can be accessed via the periphery servers known to a core server. The ScM is a Virtual Machine Monitor (VMM) \cite{Rosenblum05}; its
main function is to configure and supervise the system, rather than tightly control the virtual machines (VMs)
running on the physical server, as is the case for Xen \cite{Barham03}, Xoar \cite{Colp11}, or other VMMs.
The ScM is a component of the software stack running on each core server at all times; its main functions are:

\smallskip

\noindent 1. To respond to external events following on a set of policies and goals; e.g., choose the operation mode of the server in response to a service request and configure the server accordingly.

\smallskip

\noindent 2. To evaluate the performance of the server relative to its long- and short-term goals after the
completion of a service request. To maintain information about the environment and the past system history such as: the primary contacts, the energy consumption, the ratio of successful versus failed bids
for service, the average system utilization over a window of time, the effectiveness of the cost model used for bidding, and all other data relevant for the future behavior of the server.

\smallskip

\noindent 3. To evaluate the behavior of the application, whether the information in the service request si adequate and the resulting SLA adopted after the bid was successful.

\smallskip

\noindent 4.  To support system security and control communication with the external world.

The effectiveness of the model depends on the ability of the ScM to make most decisions
using local information thus, minimizing communication with other systems. Minimal intrusiveness is another critical requirement for the ScM; this implies that the ScM should monitor system events by observing control data structures, rather than being directly involved in the flow of control and reacting only to prevent policy violations.

\smallskip

{\bf Cloud periphery.} A self-organizing cloud includes a relatively small number of periphery  servers. The cloud periphery  plays a critical role in a self-organizing system, {\it it acts like the nervous system of the cloud} and it links the core to the outside world and with the internal services. This strategy allows the core servers to maintain a minimum amount of data about the external and the internal environment and should be dedicated to their main mission of providing services to the user community. At the same time, they should be more agile and able to respond to unforseen events, to accommodate software updates, and to balance the load placed on the auxiliary servers (AS). The main functions of a periphery server are:

\smallskip

\noindent 1. To provide an interface of the cloud with the outside world.

\smallskip

\noindent 2. To aid in self-organization, self-management, and self-repair.

\smallskip

\noindent 3. To act as a broker - sending service requests, to a subset of core servers and then forwarding bids from
core servers to the entity requesting service. Finally, they could be used to mediate the creation of a Service Level Agreement (SLA) between the two parties.

\smallskip

{\bf Auxiliary servers.} Support internal services such as: accounting, billing, policy and goal management, system statistics, patch management, software license management, and reputation and trust services.

\smallskip

{\bf Distinguishing features of the model.} Some of the mechanisms in our model have been discussed in the literature e.g., \cite{Azambuja10, Bernstein11,Brandic10,Dutreilh10,Gmach09,Lim09}, others have been incorporated in different cloud architectures \cite{Arrasjid13,Sugerman01}. We believe that enough progress has been made on a range of topics related to the model we propose; the next step is to combine existing ideas with new ones in a coherent new architecture. Several elements distinguish the model we propose from others:

\smallskip

\noindent 1. The system is reconfigurable - the autonomous computational and storage servers can operate in several modes to maximize QoS.

\smallskip

\noindent 2. Coalitions of core servers are created dynamically to respond to service requests. Winning coalitions are determined by the results of auctions.

\smallskip

\noindent 3. Virtualization simulates the interface of an object by several means including multiplexing,
aggregation, emulation, and multiplexing combined with emulation. While exiting clouds are based exclusively on multiplexing - multiple virtual machines sharing the same server - in our model, in addition to multiplexing, we consider aggregation - the creation of a single virtual server from several physical servers, and federation of servers.

\smallskip

\noindent 4. A service request is formulated in a Cloud Service Description Language (CSDL). Once a bid is selected a client-specific, legally-binding Service Level Agreement (SLA) can be negotiated between the cloud service provider and the cloud client.
\smallskip

\noindent 5. At the completion of a service request the system evaluates the performance of the service provision, the client and the application. This information can be fed back into future service negotiations.

\smallskip

We now examine the manner that the self-organization model addresses some of the challenges discussed in Section
\ref{SelfManagement}. First, it allows the three service delivery models to coexist. For example, an organization
interested in the SaaS or PaaS delivery models, could request servers operating in the M2 or M3 modes depending on expected load.  In the case of a request for IaaS the periphery would contact core servers already operating in Mode M1 or it would initiate the formation of a new coalition. The lifetime of a contract varies widely; the SLAs for a SaaS,
could be rather long, months, years, while the one for PaaS could be just for the time to complete a specific job, or for
multiple jobs extended over a period of time. QoS guarantees for systems operating in Modes M2 and M3 can be provided due to performance isolation; indeed, the leader of a coalition has access to accurate information about the resource consumption and the available capacity of the coalition members.

The inefficiencies inherent to virtualization by multiplexing cannot be eliminated; at the same time, we expect
that virtualization by aggregation will introduce other inefficiencies. Some of them will be caused by distributed
shared memory, which, depending on the interconnect, can be severe, others by the interactions between the guest
OS and distributed hypervisor.  Virtualization plays no role for coalitions of servers operating in Mode M3 and the only overhead is due to the ScM's role in shielding the group from outside actions.

\section{Virtualization by Aggregation}
\label{SSI}
\medskip

Single System Image is a computing paradigm where a number of distributed computing resources are aggregated and presented via an interface that maintains the illusion of interaction with a single system \cite{Buyya97,Buyya01,Pfister98}. The concept of seamlessly aggregating distributed computing resources is an attractive one, as it allows the presentation of a unified view of those resources, either to users or to software at higher layers of abstraction. The motivation behind this aggregation process is that by hiding the distributed nature of the resources the effort required to utilize them effectively is diminished. Aggregation can be implemented at a number of levels of abstraction, from custom hardware and distributed hypervisors through to specialized operating system kernels and user-level tools.  Single system image embodies a rich variety of techniques and implementations with a history going back over three decades. Three of these are most relevant to the self-organizing clouds concept: distributed hypervisors, kernel-level SSI and APIs that aggregate coprocessors. A brief overview of each is presented next, before the suitability of each for implementing Mode M3 is examined.

\underline{Single System Image.} There have been several implementations of distributed hypervisors that provide a single guest operating system instance with a single system image of an entire cluster \cite{Chapman05,Kaneda05,Peng08,Wang09}. This approach has the advantage of being largely transparent to the guest OS, eliminating the need for far-reaching kernel modifications and hence allowing for a wider selection of guest operating systems. These systems build a global resource information table of aggregated resources such as memory, interrupts and I/O devices. A virtualized view of these hardware resources is then presented to the guest operating system. Standard distributed shared memory techniques are used to provide a global address space. The single system image is maintained by alternating as necessary between the guest OS and the hypervisor via trap instructions.

There have been numerous implementations of kernel-level SSI, with the oldest dating back to the 1970s. There are two basic approaches in terms of design philosophy: dedicated distributed operating systems and adaptations of existing operating systems. The latter approach can be further divided into over-kernel and under-kernel implementations. Under-kernel (or \textit{``under-ware''} \cite{Walker99}) systems seek to transparently preserve existing APIs, such as POSIX. In contrast, over-kernel systems provide additional or extended APIs that allow for more efficient utilization of distributed resources. The under-kernel approach is of most interest here as it allows existing application code to run without modification.

Notable implementations of kernel-level SSI include MOSIX \cite{Barak12}, OpenSSI \cite{Walker03} and Kerrighed \cite{Morin04}. Of these, Kerrighed provides the greatest level of transparency, supporting a unified file system, process space (including migration) and memory space. Uniquely for a Linux-based system, Kerrighed’s distributed memory model allows for thread migration, albeit with the attendant inefficiencies caused by OS-managed remote paging; early experiments on commodity hardware resulted in a significant slowdown compared to a true SMP machine \cite{Margery03}. The extent of the transparency provided by Kerrighed is highlighted by an experiment where detailed load balancing results for Kerrighed could not be obtained because it was not possible to determine the load on individual cluster nodes as commands such as \texttt{ps} display all processes running cluster-wide \cite{Osinski09}.

A number of abstraction models have been proposed for co-processors that present a single system image to the application developer or runtime environment. For example, RASCAL (Reconfigurable Application-Specific Computing Abstraction Layer) is a software library developed by SGI that provides functionality for device allocation, reconfiguration, data transport, error handling and automatic wide scaling across multiple FPGAs \cite{Cofer08}. Virtual OpenCL (VCL) \cite{Barak11} provides a similar API-level abstraction for GPUs distributed across a cluster. Depending on the size of the request, the VCL runtime environment may allocate one or more GPUs installed in a single machine or a collection of devices spanning several machines. The Symmetric Communication Interface (SCIF) API performs a similar role for Intel Many-Integrated-Core (MIC) coprocessors \cite{Intel13}.

\underline{Cloud resource aggregation.} Mode M3 of the self-organizing cloud architecture is characterized by the aggregation of physical computational resources into virtual servers. At this speculative stage of development, it is too early to definitively identify the most appropriate aggregation mechanisms. Nevertheless, a review of the strengths and weaknesses of candidate techniques and technologies is instructive.

Creating virtual cloud servers by abstracting at the hypervisor level has the compelling advantage that existing virtual machine images, operating systems, and application code can run unmodified. Under this approach, a self-configuration monitor would examine the role assigned to it as part of a bid and determine if it needs to scale horizontally. If so, it pulls in other servers which merge with it to create an aggregated virtual server using a distributed hypervisor. As with all distributed shared memory systems, the performance of the overall system is closely bound with that of the interconnect used, and applications with unsuitable memory access patterns can undergo significant performance degradation compared to a true SMP machine. For those applications that are suitable, however, the transparent aggregation would allow for a significant degree of horizontal scaling without any changes to virtual machine image, operating system or application code. A drawback of the hypervisor-level approach is the inability to scale the virtual server up or down after it has been created given the current state of distributed hypervisor and operating system functionality.

The operating system approach to aggregation would be similar to that for hypervisors: in response to a bid, a number of virtual machine instances, potentially across a number of physical machines. These virtual machine instances would run a single system image operating system, such as Kerrighed, and as such could be combined into clusters, each of which would aggregate the underlying virtual resources in a transparent fashion. The benefits of this approach are that using higher-level operating system abstractions, such as processes and file systems, may result in more efficient distributed execution of some workloads compared to the low-level shared memory approach taken by distributed hypervisors. Furthermore, advanced SSI implementations such as Kerrighed allow nodes to join and leave the cluster dynamically, facilitating elastic scaling. The most obvious drawback is the need to use a specific SSI operating system, which may lead to software compatibility issues. More fundamentally, the design assumptions that underpin software packages may preclude the SSI approach. For example, a scalable web serving cluster might utilize multiple instances of the Apache web server running on multiple physical or virtual machines, which are clustered together to load balance incoming requests. Replicating this arrangement using process migration would be difficult, as a host of issues, such as filesystem contention, would arise if multiple instances of Apache were run simultaneously \cite{Pfister09}.

The various single system image abstraction of distributed coprocessing resources could also have a role in self-optimizing clouds. Several cloud providers support the creation of VM instances with attached GPU coprocessors, with support for FPGAs and MICs likely to become available in the future. This heterogeneity of compute resources would provide self-organizing clouds with a rich palette in terms of how bids could be configured, at the cost of a much larger number of potential resource combinations. One approach would be to statically associate coprocessing resources with virtual machines when responding to bids. So, if a VM is expected to run an application that can benefit from GPU acceleration then a suitable number of GPU cores can be assigned to the VM as a single virtual GPU. The obvious drawbacks to this approach are efficiency and scalability. Granting virtual machines exclusive access to coprocessors prevents those coprocessors being used by other VMs when there is no work available. Similarly, static coprocessor allocations limit the horizontal scalability available to individual VMs by preventing them from taking advantage of idle coprocessors in the resource pool that are not assigned to them. The ideal solution, therefore, is API support at the application level so that the global resource pool of coprocessors can be utilized by application running in VMs. However, this approach would require support at the application level.

\section{Simulation Experiments}
\label{Simulation}
\medskip
      

\begin{figure*}[!ht]
\begin{center}
\includegraphics[width=14.5cm]{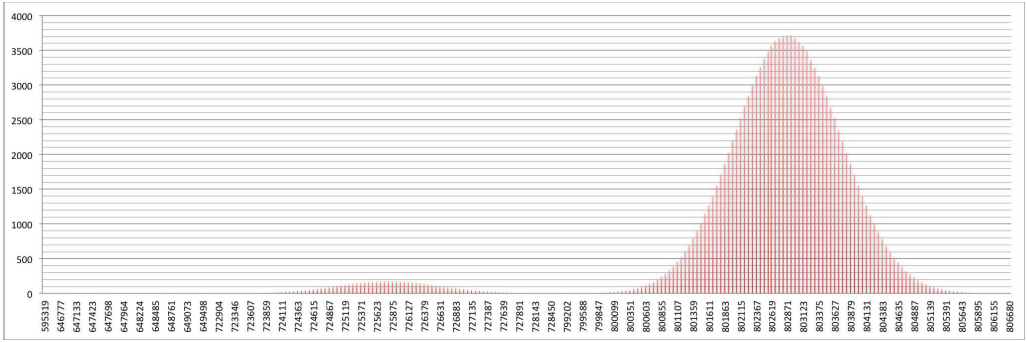}
\end{center}
\caption{Histogram of the number of secondary contacts for $m=10$.}
\label{SecondaryContactsFig}
\end{figure*}

It is impractical to experiment with a very large number of physical systems, so we chose numerical simulation to investigate the feasibility of some of the ideas discussed in this paper. We believe, that at this stage, the emphasis should be on qualitative rather than quantitative results. Thus, our main concern when choosing the parameters of the simulation was to reflect the scale of the system and  ``typical'' situations.  The very large number of core servers we chose to experiment with forced us to consider simpler versions of the protocols for bidding.

The simulation experiments reported in this section run on the Amazon cloud. Several storage optimized  {\tt hi1.4xlarge} instances\footnote{An {\tt hi1.4xlarge} instance can deliver more than $120,000$ $4$ KB random read IOPS and between $10,000$ and $85,000$ $4$ KB random write IOPS. The maximum throughput is approximately $2$ GB/sec read and $1.1$ GB/sec write.} running on a 64-bit architecture were used.
This choice was motivated by the scale of the system we simulated; the description of the very large number of core servers requires a very large address space thus, systems with access to a very large physical memory.  The simulation ran concurrently on $16$ virtual cores (vCPUs) delivering 35 ECUs (Elastic Cloud Units) and used $60.5$ GB of main memory. The instances were launched on the same cluster and servers were connected by a non-blocking $10$ Gbps Ethernet. The simulation, implemented in {\it C++}, used extensively multi-threading to reduce the execution time; each one of the 16 virtual cores ran a single thread.

First, we report on experiments for the initial self-organization stage.  The  self-organization algorithm consists of the following steps: (i) the $M$ periphery servers broadcast their identity and address; (ii) a core server randomly selects $m$ of the $M$ periphery servers, includes them in its {\it cps} list, and informs each one of them about its selection; (iii) a periphery server constructs the list of core servers known to it, {\it pcs}.

We chose several parameters to configure the system : (a) the number $N$=8,388,608 of core servers was randomly chosen  in the $10^{6} - 10^{7}$ range; (b) the number of periphery servers was chosen to be $M=1,000$; and (c) the size $n=500$ of the primary contact list, {\it pcl}, maintained by each core server. To determine the number of secondary contacts for the $N$ core servers, we experimented with several values for the size of {\it pl}, the list of periphery servers kinown to a core server,  $m=5, 10, 20, 50$.

We study $p$, the size of the {\it pcs} list maintained by each one of the $M$ periphery servers; this list gives the number of the core servers known to each periphery server.  As expected, $\bar{p}$, the average $p$, increases  when $m$ increases, the dependence is nearly linear:

\begin{equation}
\begin{array}{ccc}
m  & \bar{p} & \bar{p}/N \\
5  & 41,859  &  0.494 \% \\
10 & 83,593  &  0.996 \% \\
20 & 166,363 &  1.983 \% \\
50 & 409,707 &  4.884 \% \\
\end{array}
\end{equation}

We are also interested in the distribution of the number of  secondary contacts $S$ that can be reached by each one of the $N$ core servers when required. The histogram of the distribution of the secondary contacts, shows that when $m=5$ some $7,200$ core servers have a number of secondary contacts close to $207,423$, the average number of secondary contacts. On the other hand, when $m=50$ we have a multi-modal distribution; for example, there are some $1,500$ core servers claiming $93\%$ of the total number of core servers as a secondary contact. When $m=50$ each core server is able to claim a very large fraction of all other core servers, and there are $8,388,608$ of them, as secondary contacts. The average number of secondary contacts, $\bar{S}$, increases with  $m$ but much faster than $\bar{p}$:

\begin{equation}
\begin{array} {ccccc}
 m & \min(S) & \max(S) & \bar{S} &  \bar{S}/N      \\
5  &   124,303 &   210,086 &   207,423 & 2.472 \%  \\
10 &   490,466 &   807,182 &   799,403 & 9.529 \%  \\
20 & 2,289,093 & 2,797,363 & 2,768,772 & 33.006 \% \\
50 & 7,241,339 & 7,748,921 & 7,701,688 & 91.811 \% \\
\end{array}
\end{equation}

\begin{figure*}[!ht]
\begin{center}
\includegraphics[width=9.4cm]{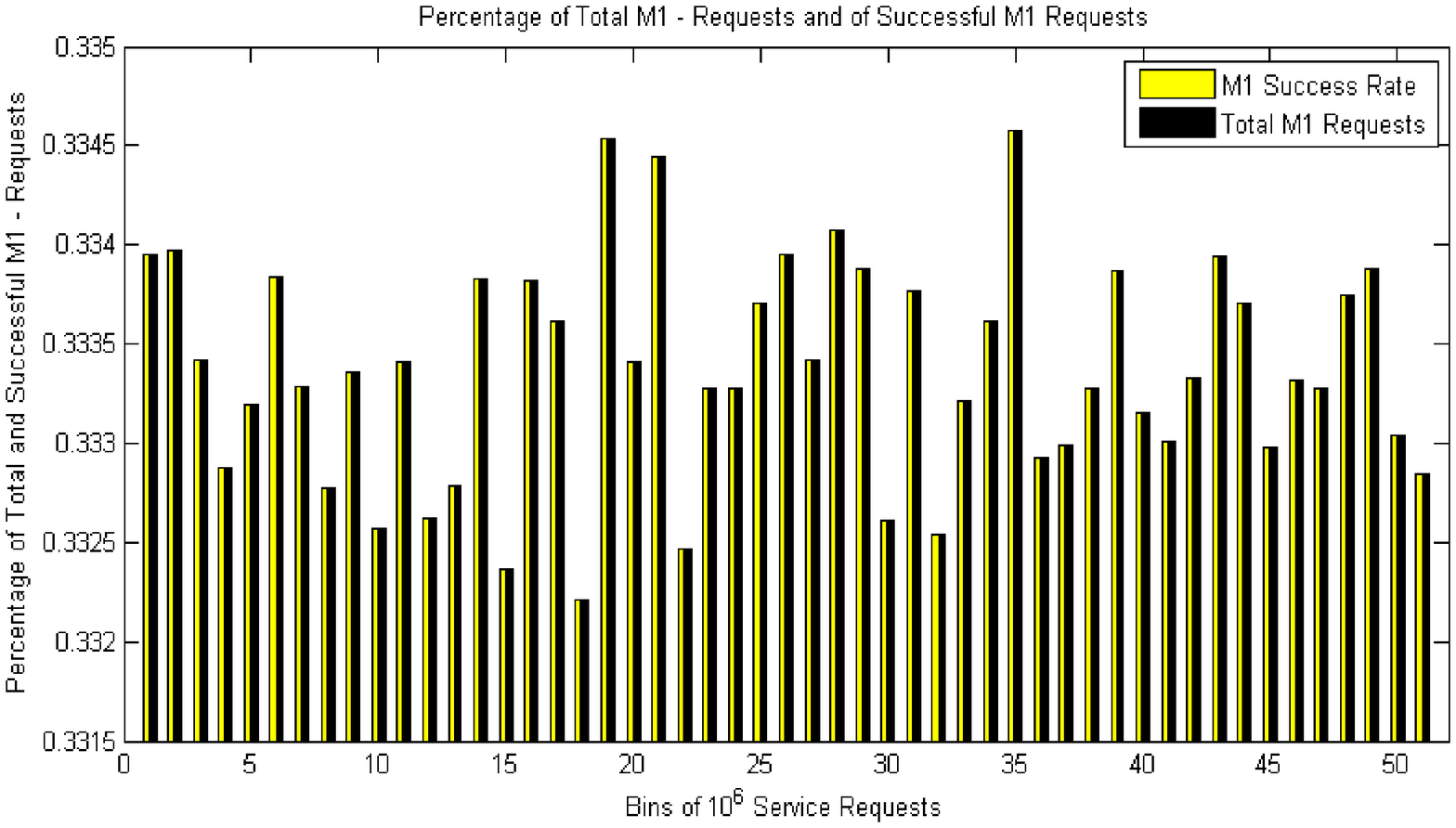}
\includegraphics[width=9.4cm]{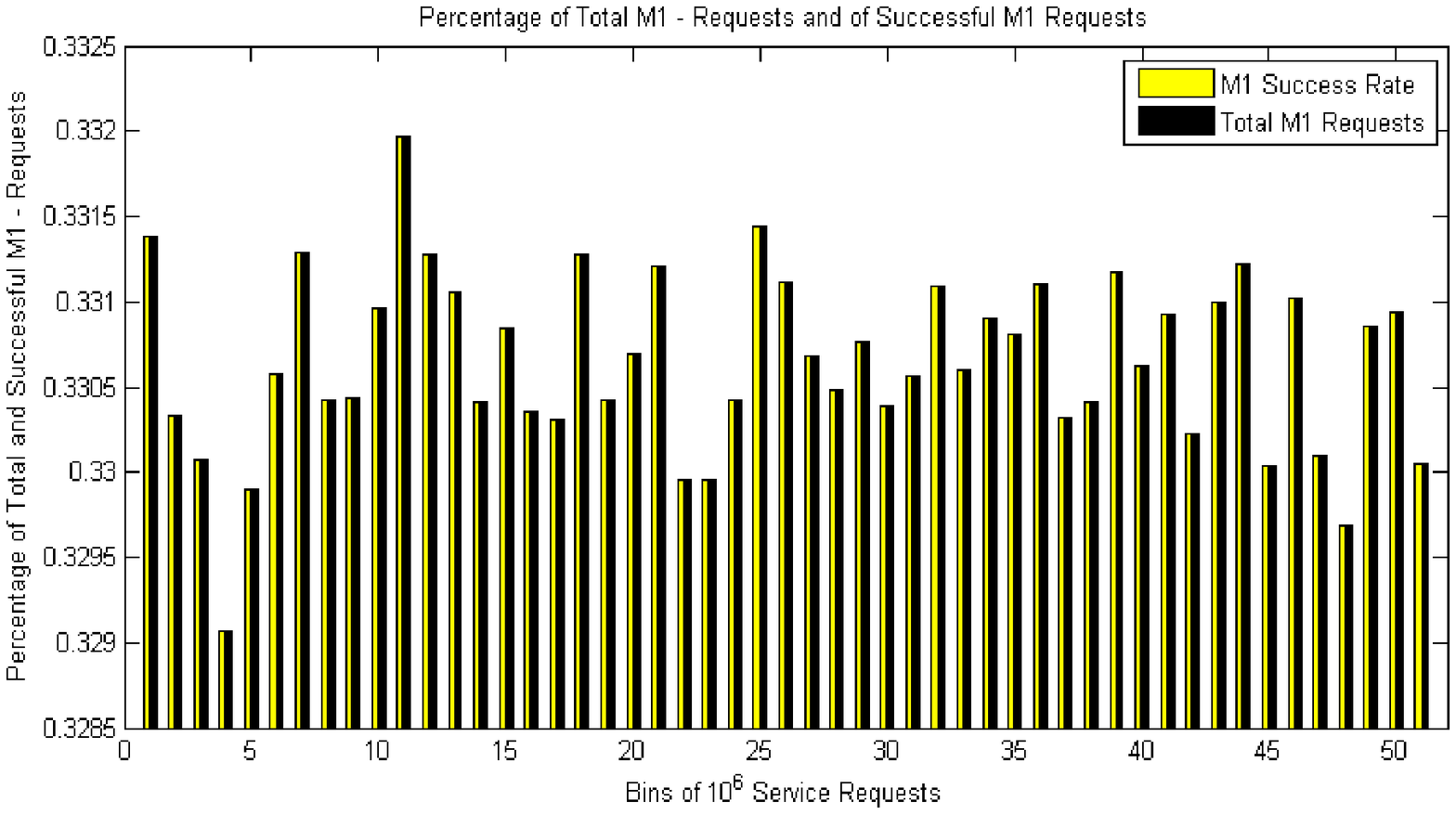}\\
(a)~~~~~~~~~~~~~~~~~~~~~~~~~~~~~~~~~~~~~~~~~~~~~~~~~~~~~~~~~~~~~~~~~~~~~~~~~~~~~~~(b)~~\\
\includegraphics[width=9.4cm]{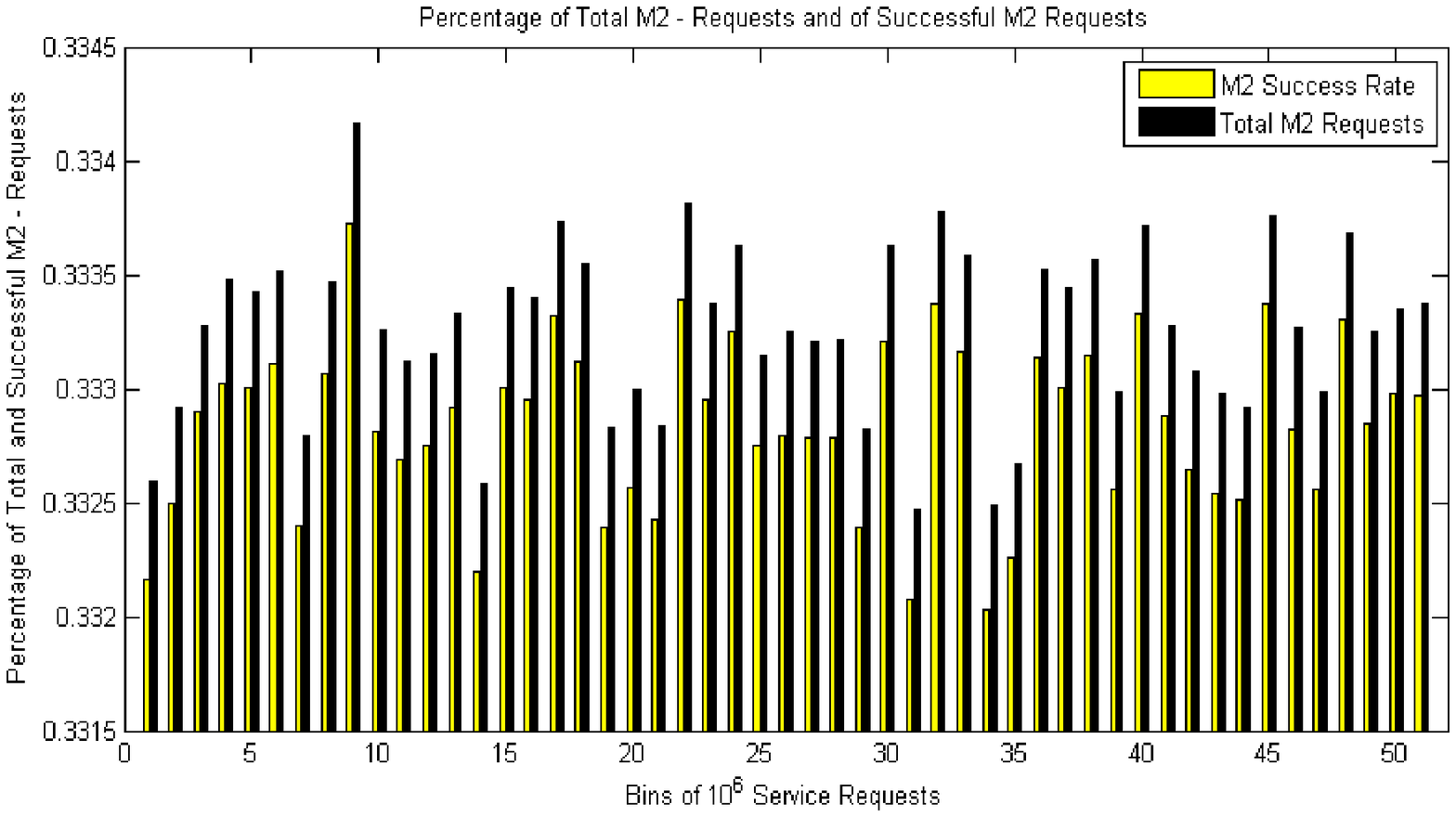}
\includegraphics[width=9.4cm]{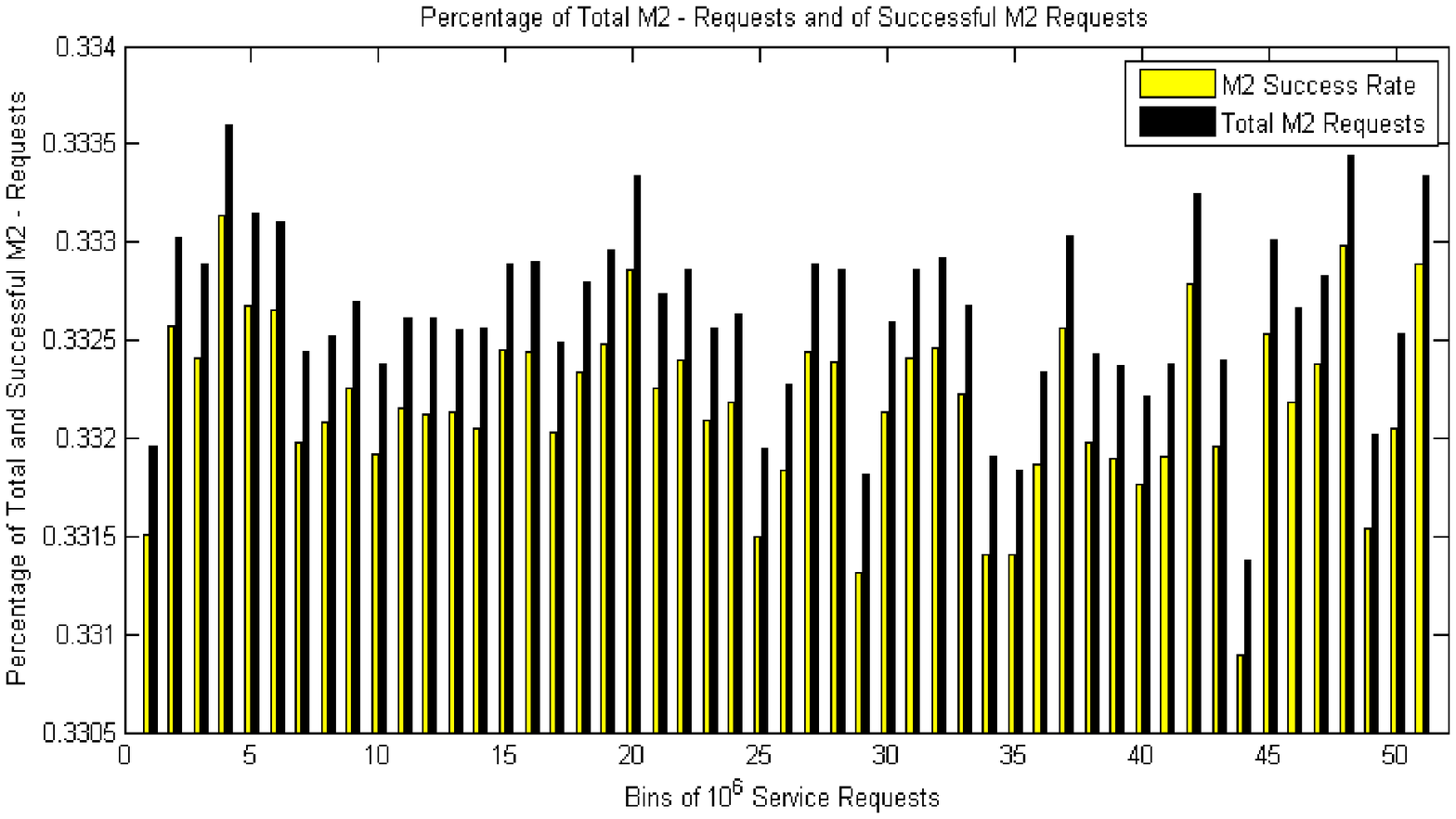}\\
(c)~~~~~~~~~~~~~~~~~~~~~~~~~~~~~~~~~~~~~~~~~~~~~~~~~~~~~~~~~~~~~~~~~~~~~~~~~~~~~~~(d)~~\\
\includegraphics[width=9.4cm]{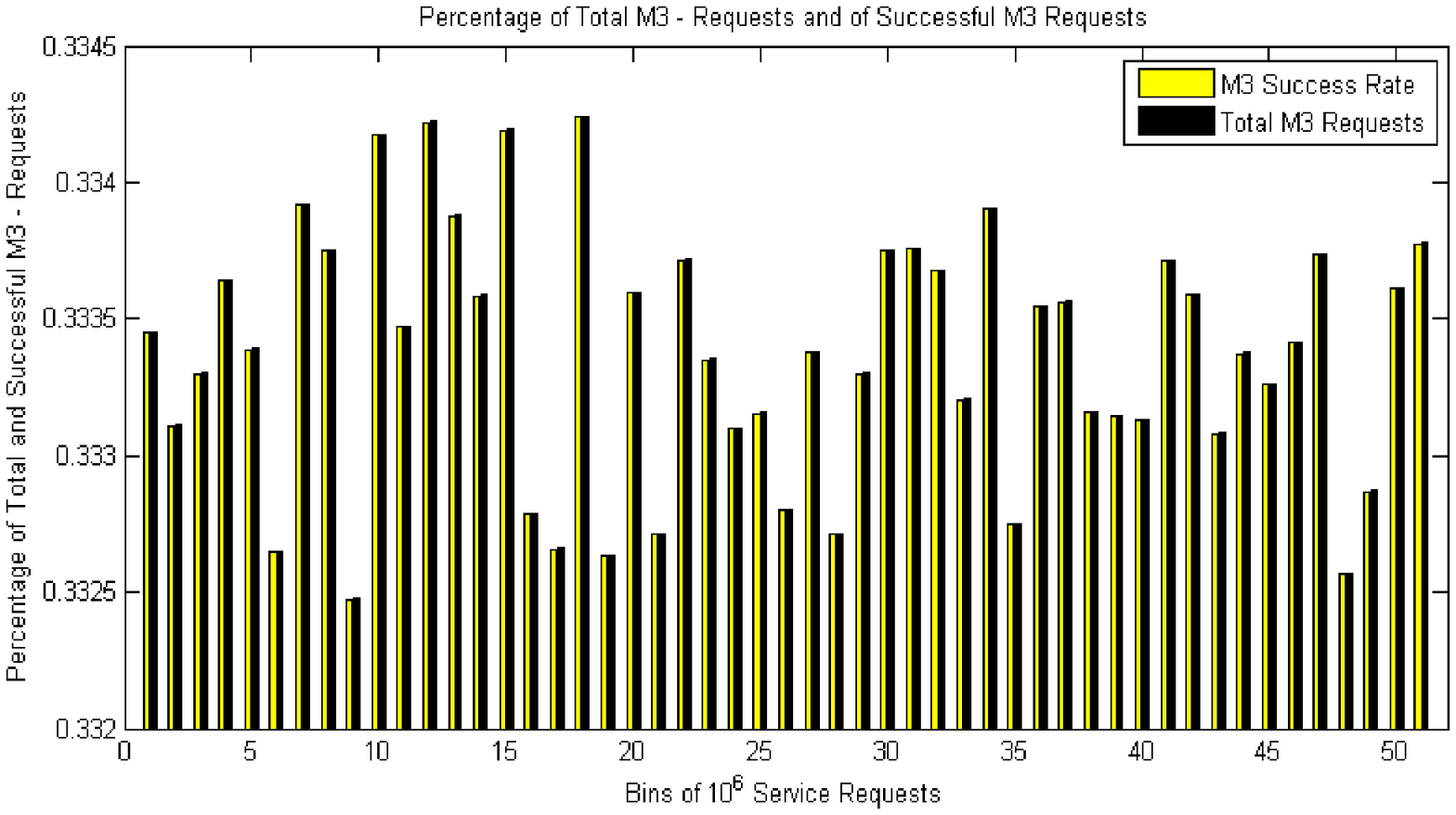}
\includegraphics[width=9.4cm]{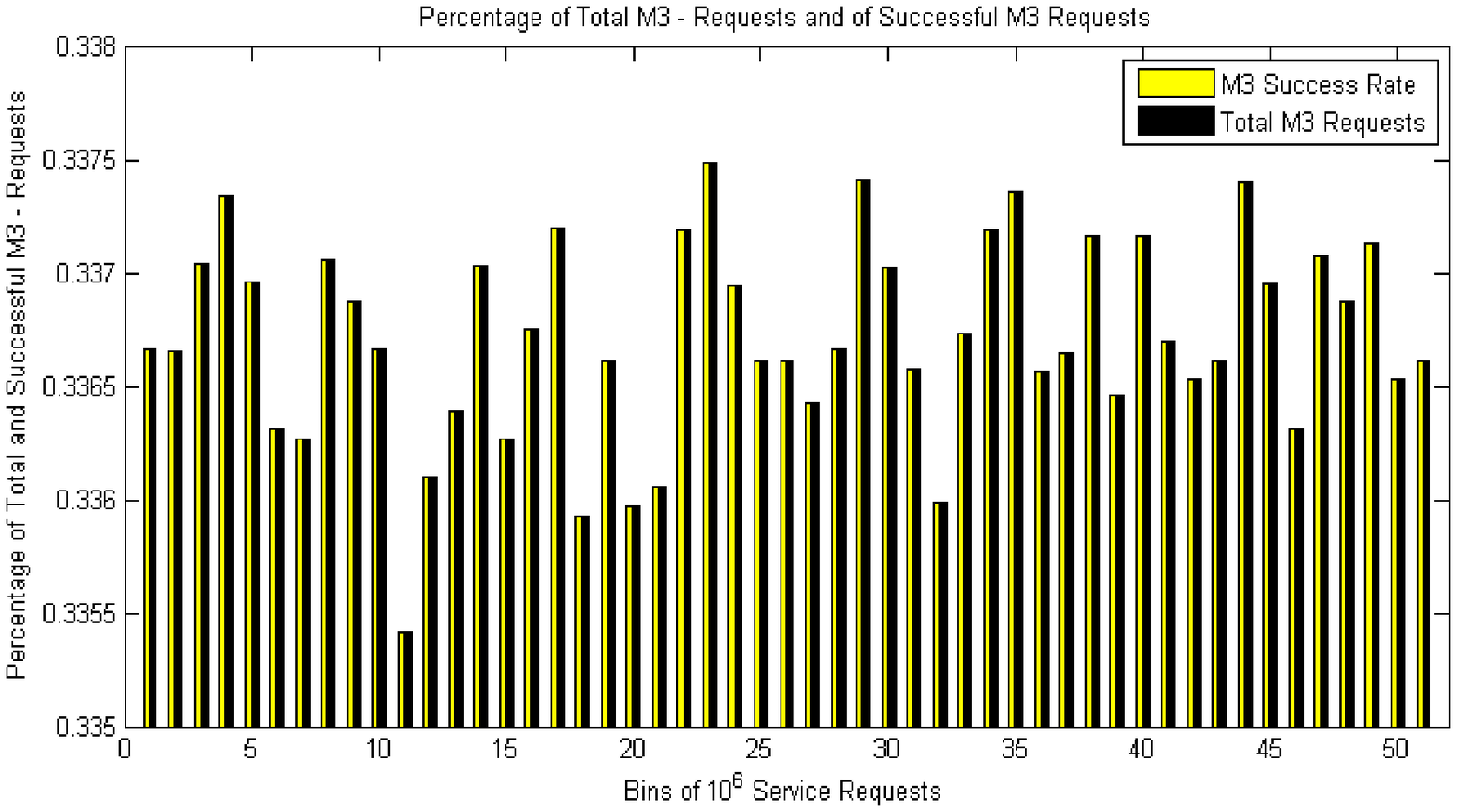}
(e)~~~~~~~~~~~~~~~~~~~~~~~~~~~~~~~~~~~~~~~~~~~~~~~~~~~~~~~~~~~~~~~~~~~~~~~~~~~~~~~(f)
\end{center}
\caption{The ratio of requests and the success rate: (a,b) M1 mode, (c,d) M2 mode, and (e,f) M3 mode. (a,c,e) {\it Exp1}, (b,d,e) {\it Exp2}.}
\label{LowLoadFig}
\end{figure*}

\begin{table*}[!ht]
\begin{center}
\caption{Summary of results for experiments {\it Exp1} and {\it Exp2}.}
\label{LowLoadTable}
\begin{tabular} {|l|l|l|l|l|l|}
\hline
Mode & Exp & Requests & Percent of total   &  Failed   & Success rate  \\
     &      & in a bin & $\times 10^{6}$   &  requests &   ( $\%$)     \\    \hline \hline
M1   & {\it Exp 1} & 17,001,928  & 0.33370 &   0       &   100  \\
     & {\it Exp 2} & 16,871,479  & 0.33049 &   0       &   100  \\ \hline
M2   & {\it Exp 1} & 16,974,809  & 0.33260 &  21,254   &   99.875  \\
     & {\it Exp 2} & 16,972,653  & 0.33049 &  23,630   &   99.61 \\ \hline
M3   & {\it Exp 1} & 17,001,928  & 0.33370 &   0       &   100  \\
     & {\it Exp 2} & 17,180,818  & 0.33655 &   0       &   100  \\ \hline
\end{tabular}
\end{center}
\end{table*}

\begin{figure*}[!ht]
\begin{center}
\includegraphics[width=9.2cm]{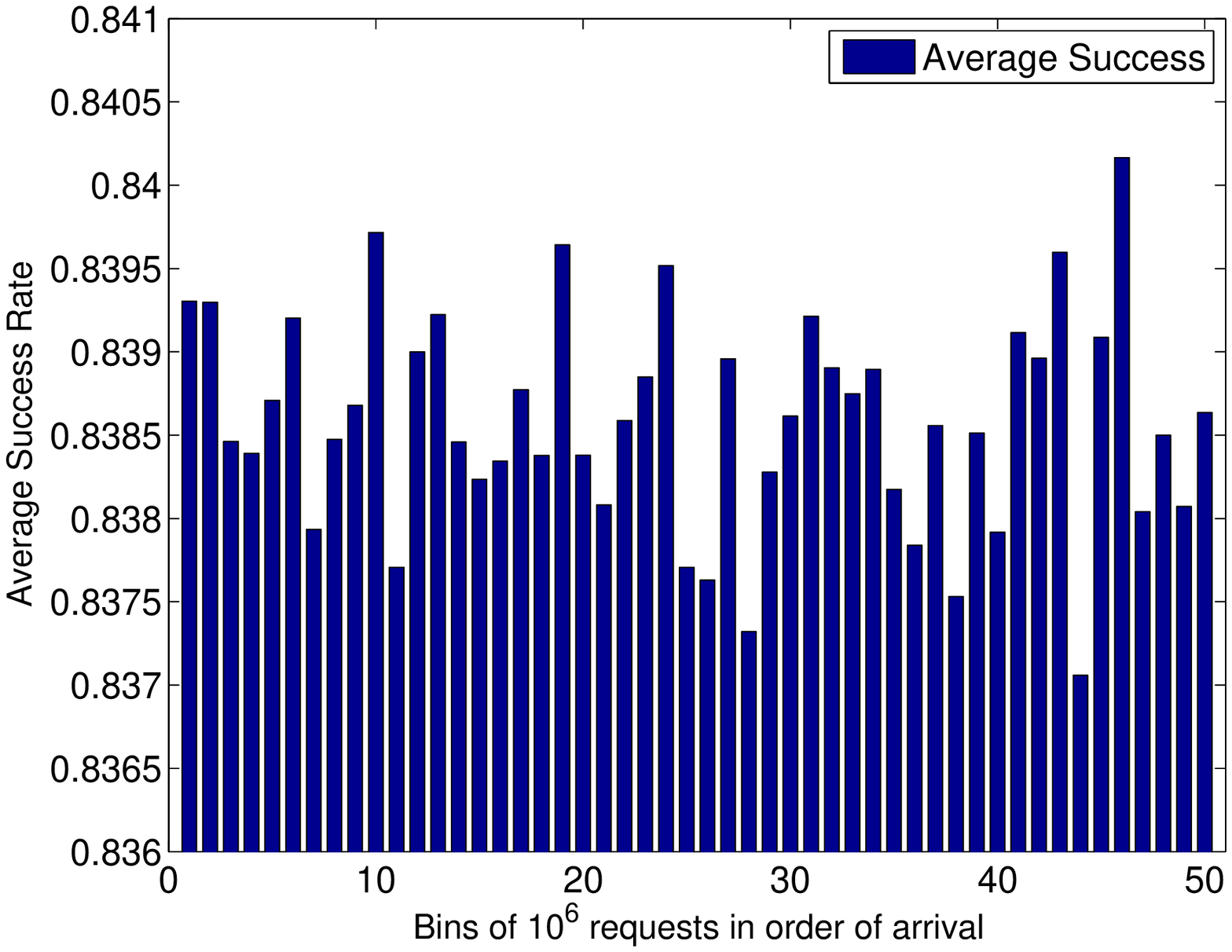}
\includegraphics[width=9.2cm]{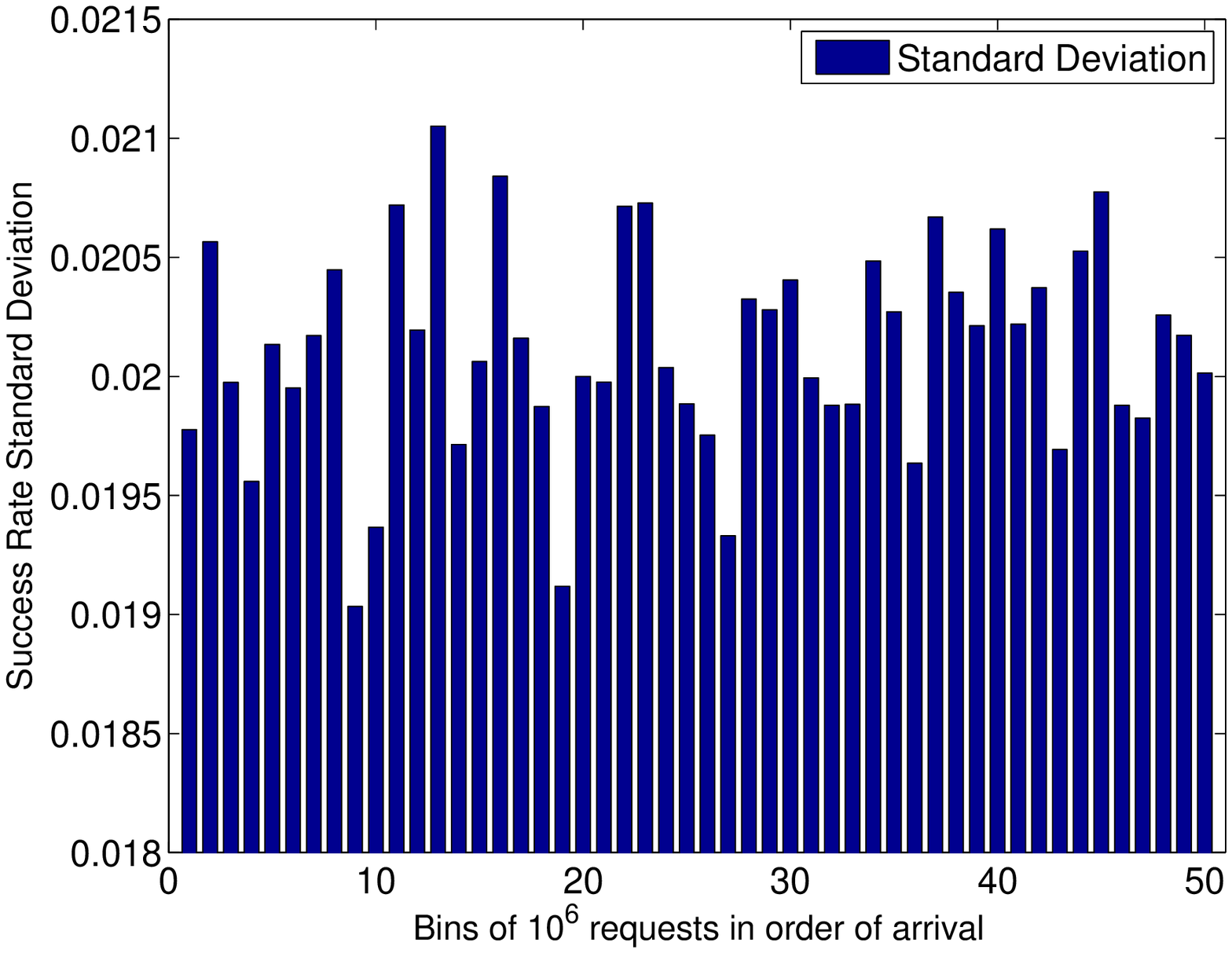} \\
\includegraphics[width=9.2cm]{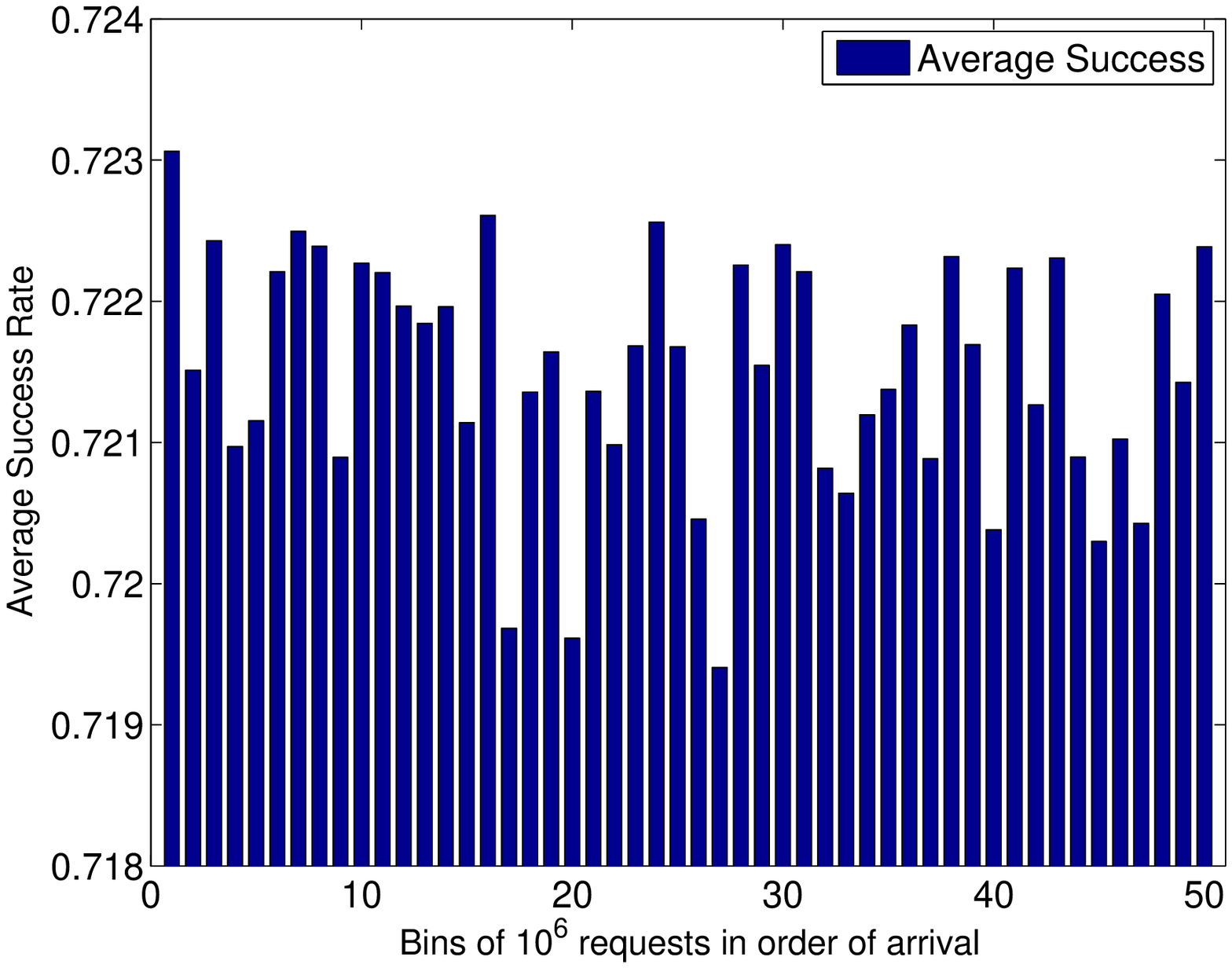}
\includegraphics[width=9.2cm]{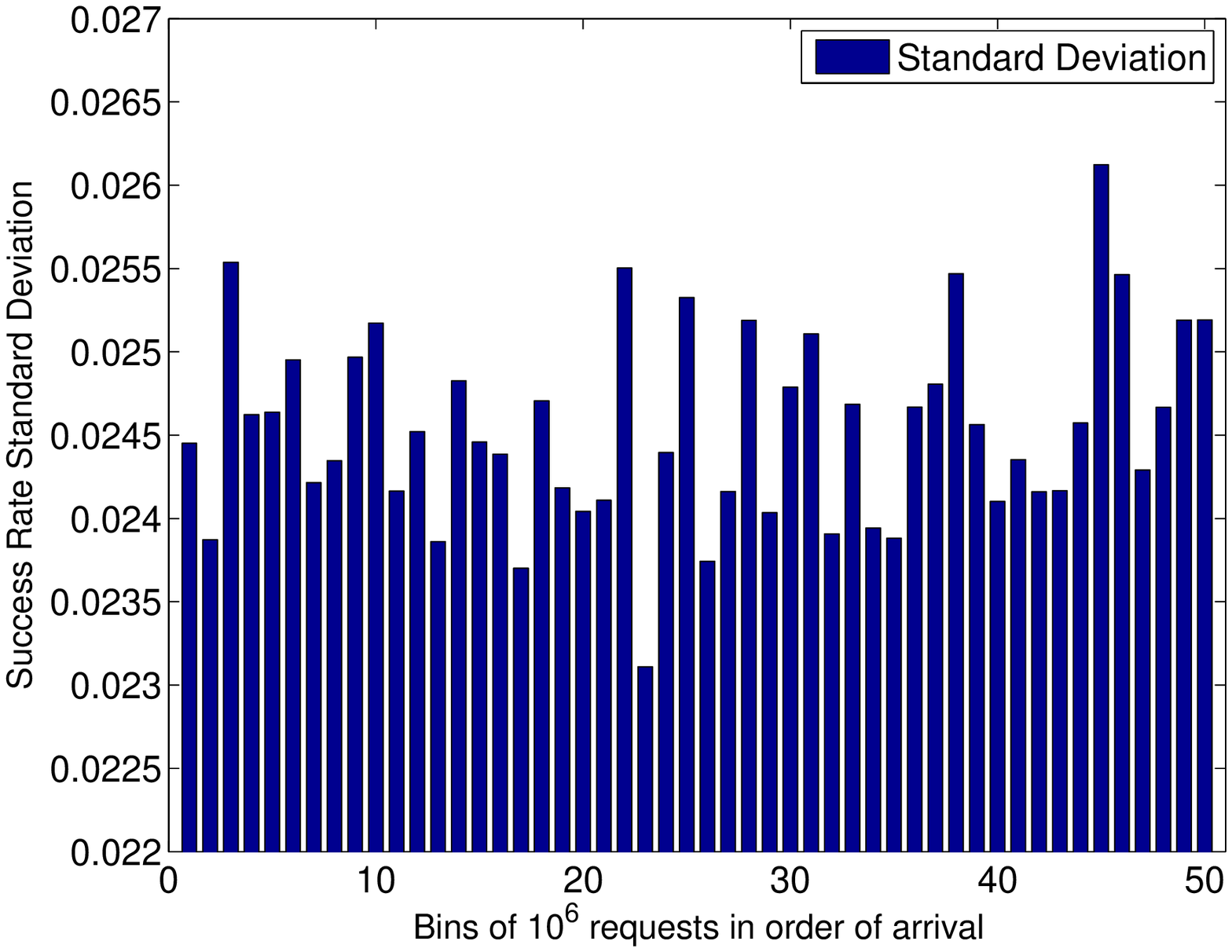}\\
\includegraphics[width=9.2cm]{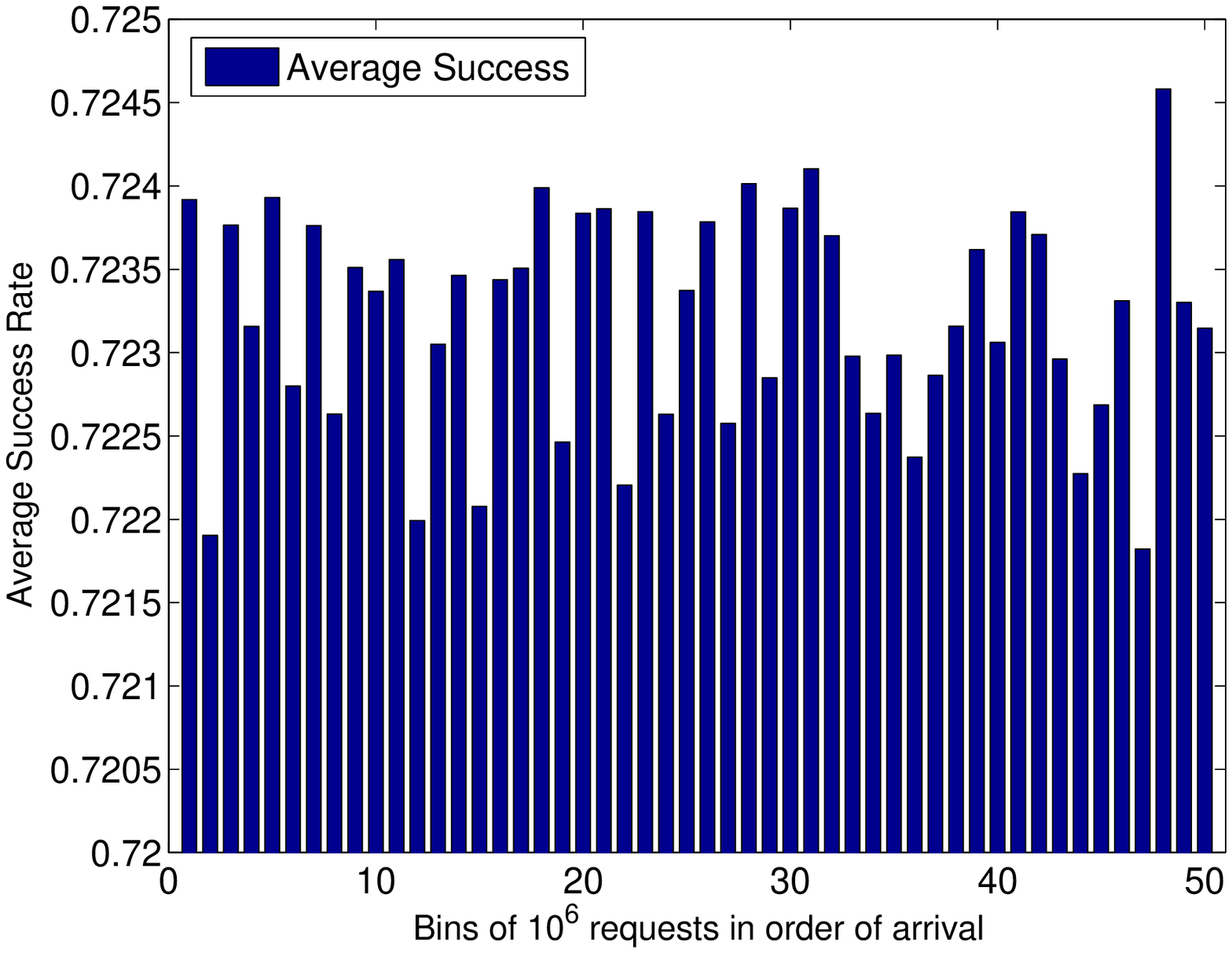}
\includegraphics[width=9.2cm]{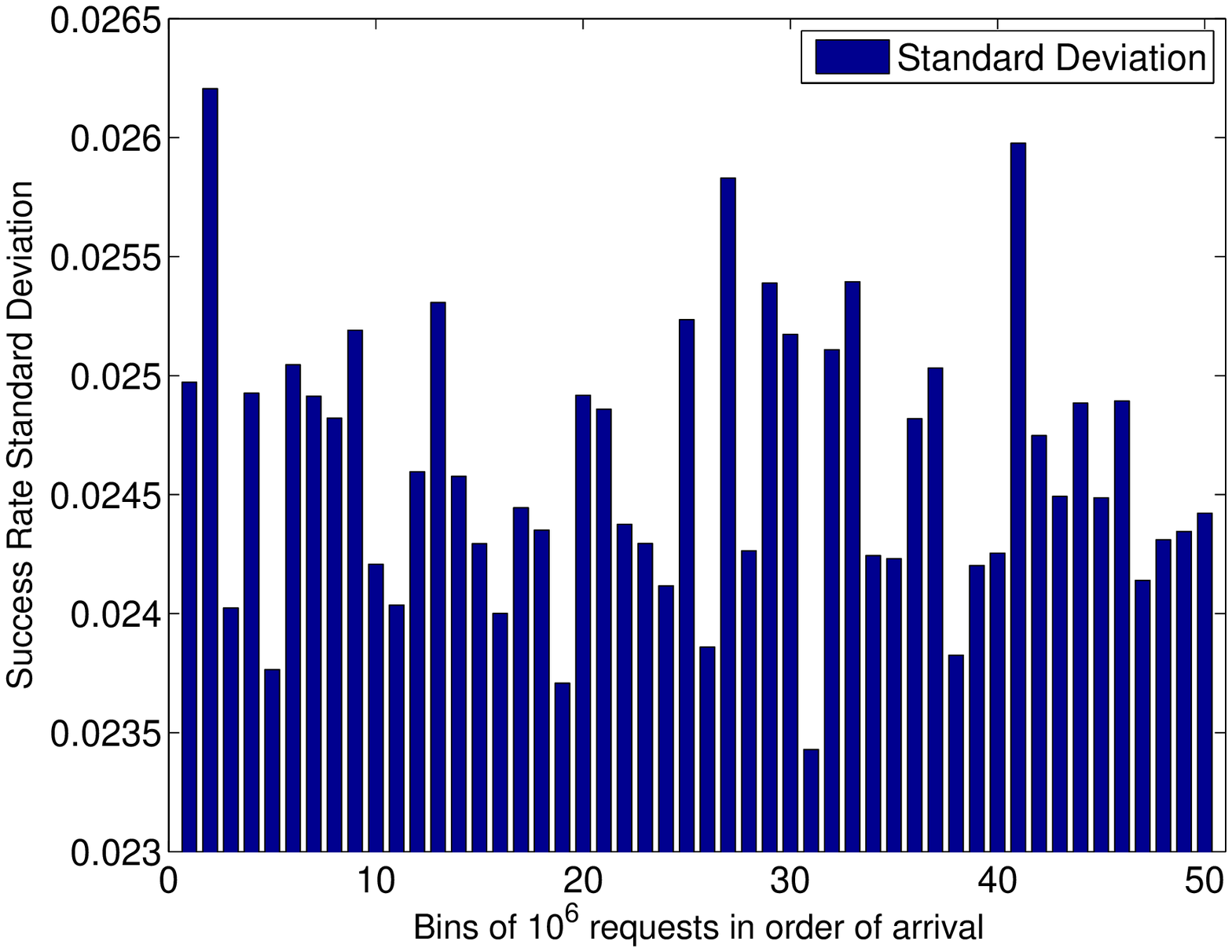}
\end{center}
\caption{Average and standard deviation of the success rate for M1, M2,M3 service requests. Coalitions initiated by a core server, {\it Exp3}.}
\label{SuccessRateCore}
\end{figure*}

Based on these results we concluded that $m=10$ is a good choice for the number of periphery servers known to each core server, see Figure \ref{SecondaryContactsFig}. This choice provides each core server with a sufficient  number of potential secondary contacts - about $10\%$ of the total number of core servers.

The execution time for the self-organization stage was about two hours.

\medskip

The second group of experiments is designed to show if the auctions, central to the architecture we propose, enables the system to respond effectively to outside requests. There are two modes to create a coalition:

\begin{itemize}

\item
C1. Coalitions are initiated by a periphery server. After receiving a service request a periphery server  sends an invitation to a subset of the core servers in its pcs list to join a coalition. The message invites bids and includes the details of the service request. When one or more coalitions is formed, a leader is elected and the periphery server transmits the bid(s) to the client. Finally the periphery server may negotiate the generation of an SLA.

\item
C2. Coalitions are initiated by a core server. After receiving a service request, a periphery server  sends an invitation to a subset of core the servers known to it and provides the details of the service request. The Zookeeper ({\it http://zookeeper.apache.org/}) coordination software based on the Paxos algorithm \cite{Lamport98,Lamport01} can then be used to elect a leader among the candidates. The leader then uses its primary contacts and if necessary its secondary contacts to build a coalition. If successful it generates a bid. In our experiments a periphery server selects a subset of $0.1\%$ of the core servers connected to it and invites them to become a leader.

\end{itemize}

The first mode is likely to require less communication overhead thus, shorter time to generate a coalition, but also considerably more intensive participation of the periphery servers which could then become a bottleneck of the system. The second approach seems more aligned to self-organization principles thus, more robust, as the periphery servers play only the role of an intermediary.

\begin{figure*}[!ht]
\begin{center}
\includegraphics[width=9.2cm]{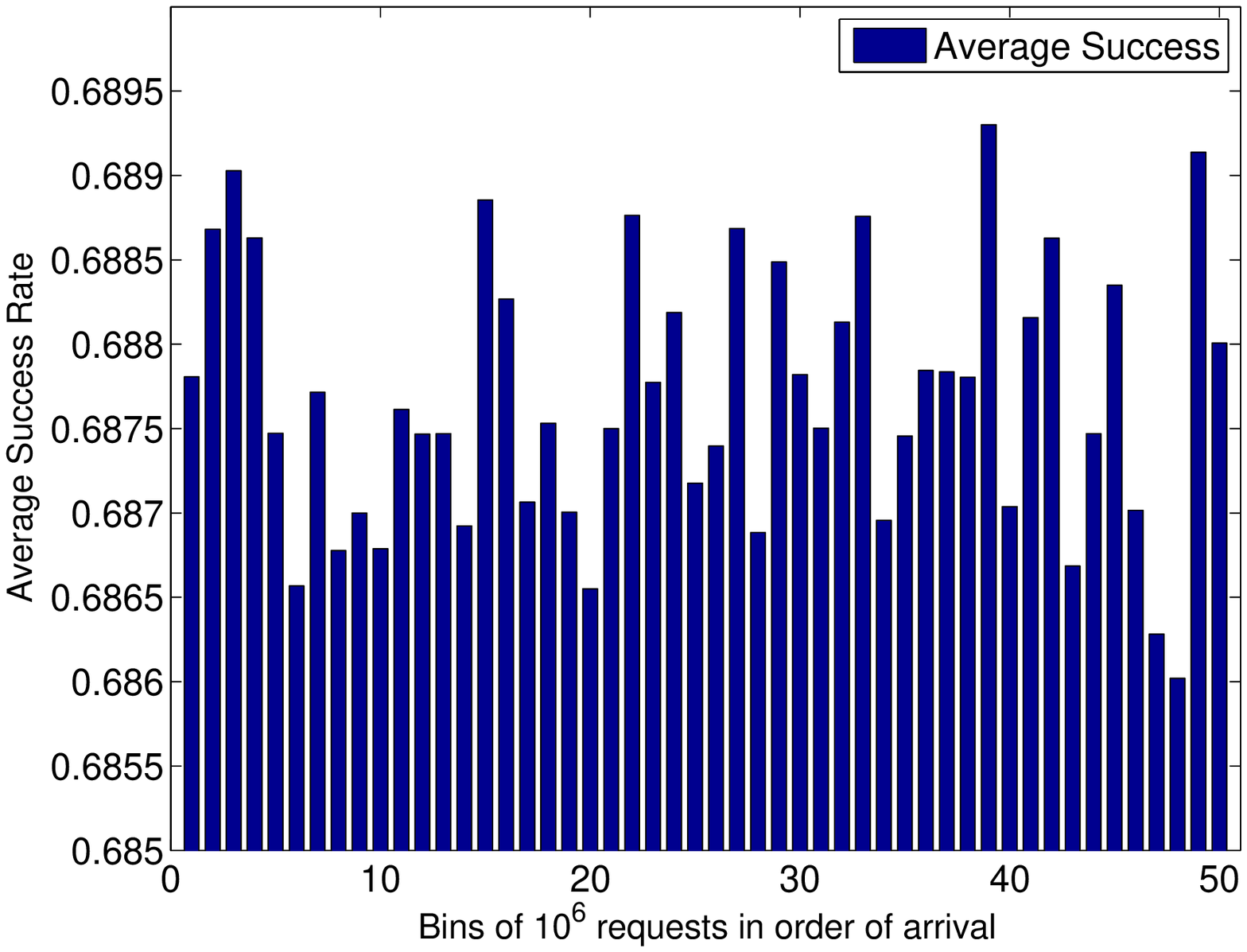}
\includegraphics[width=9.2cm]{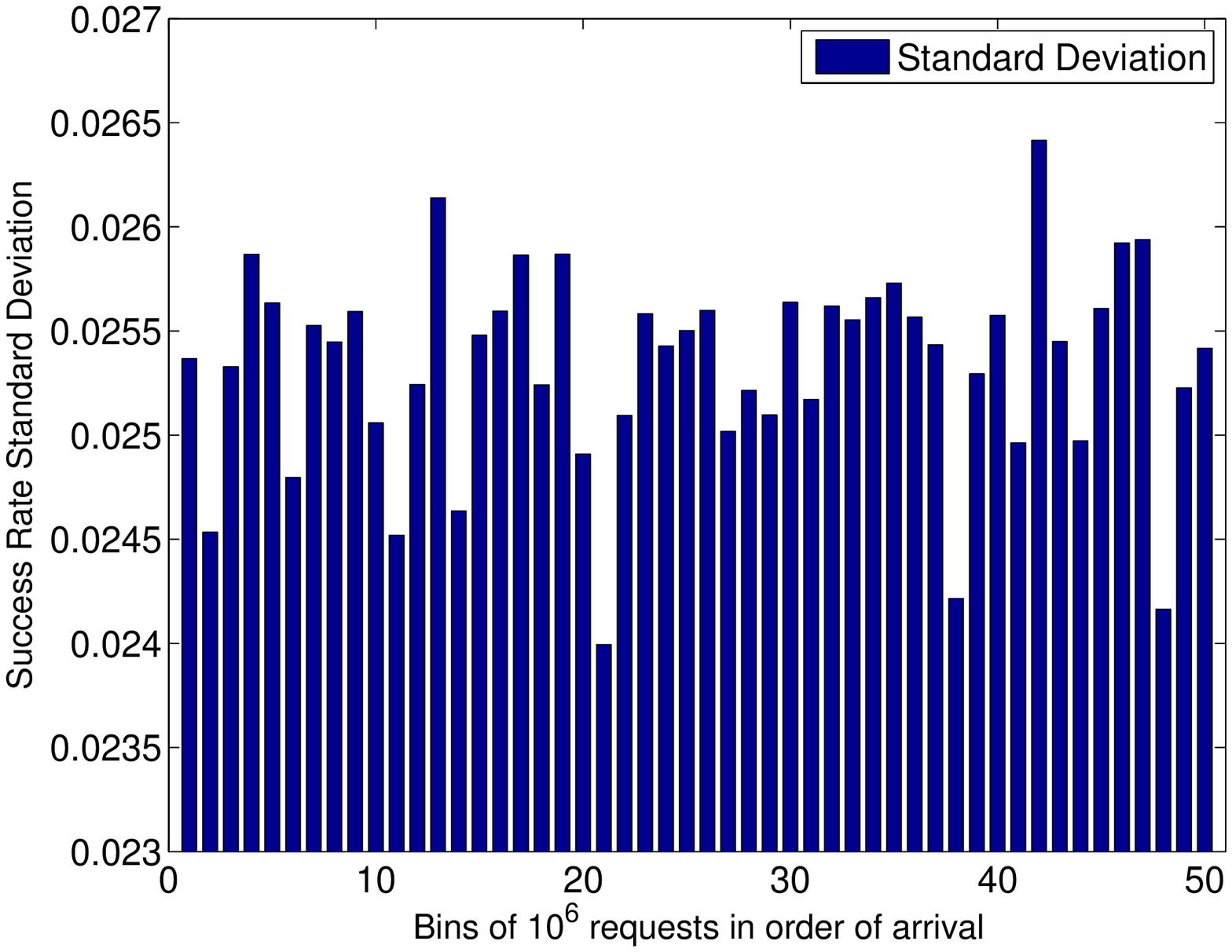} \\
\includegraphics[width=9.2cm]{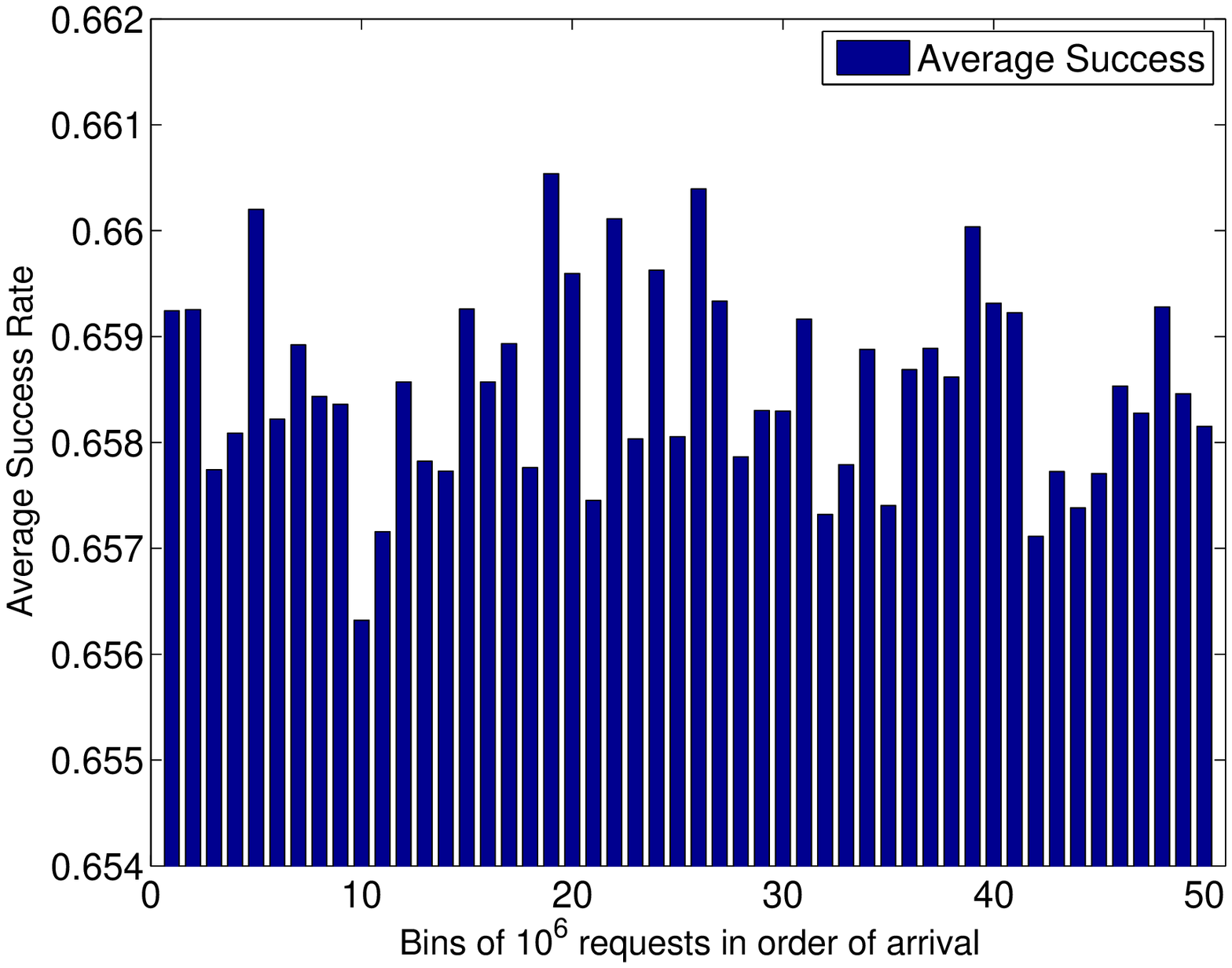}
\includegraphics[width=9.2cm]{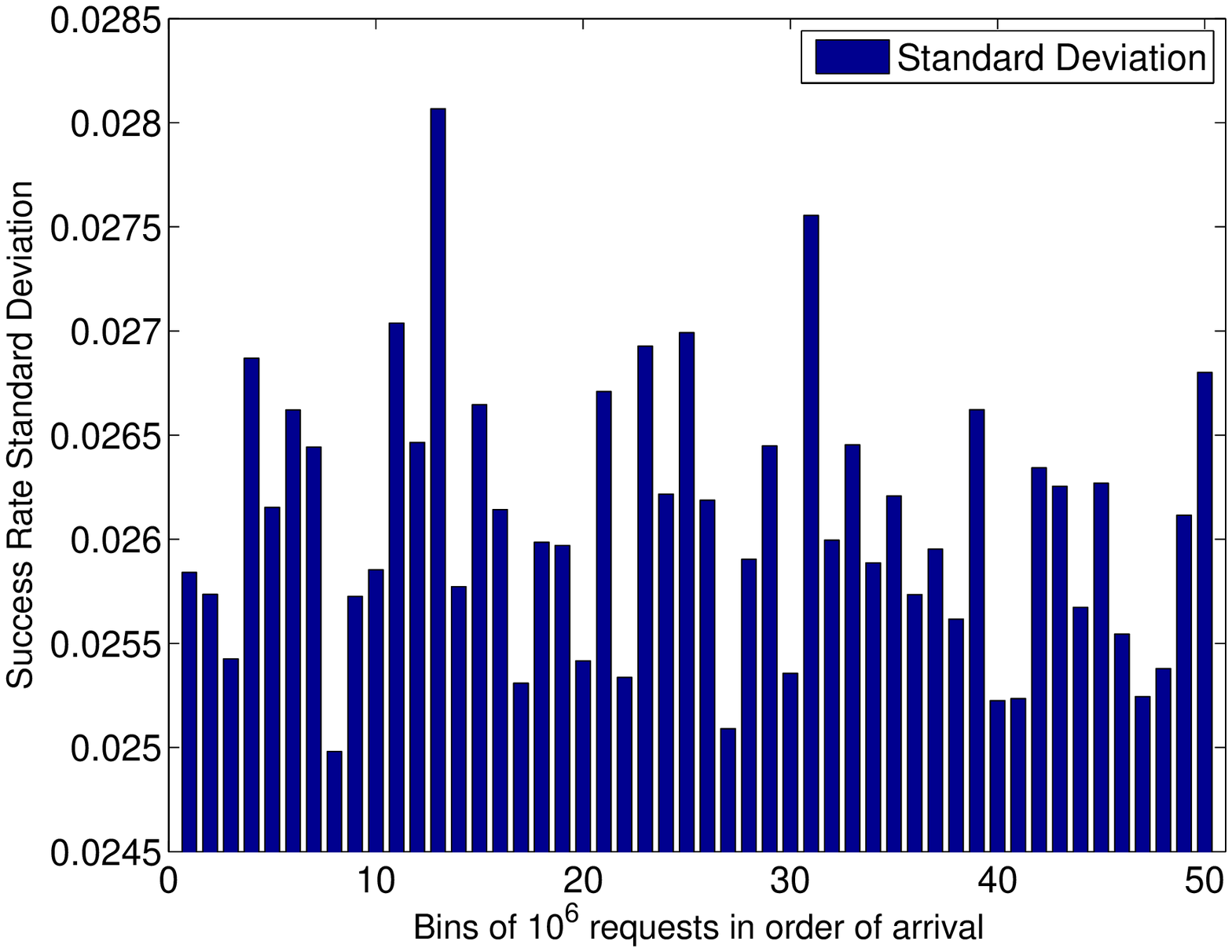}\\
\includegraphics[width=9.2cm]{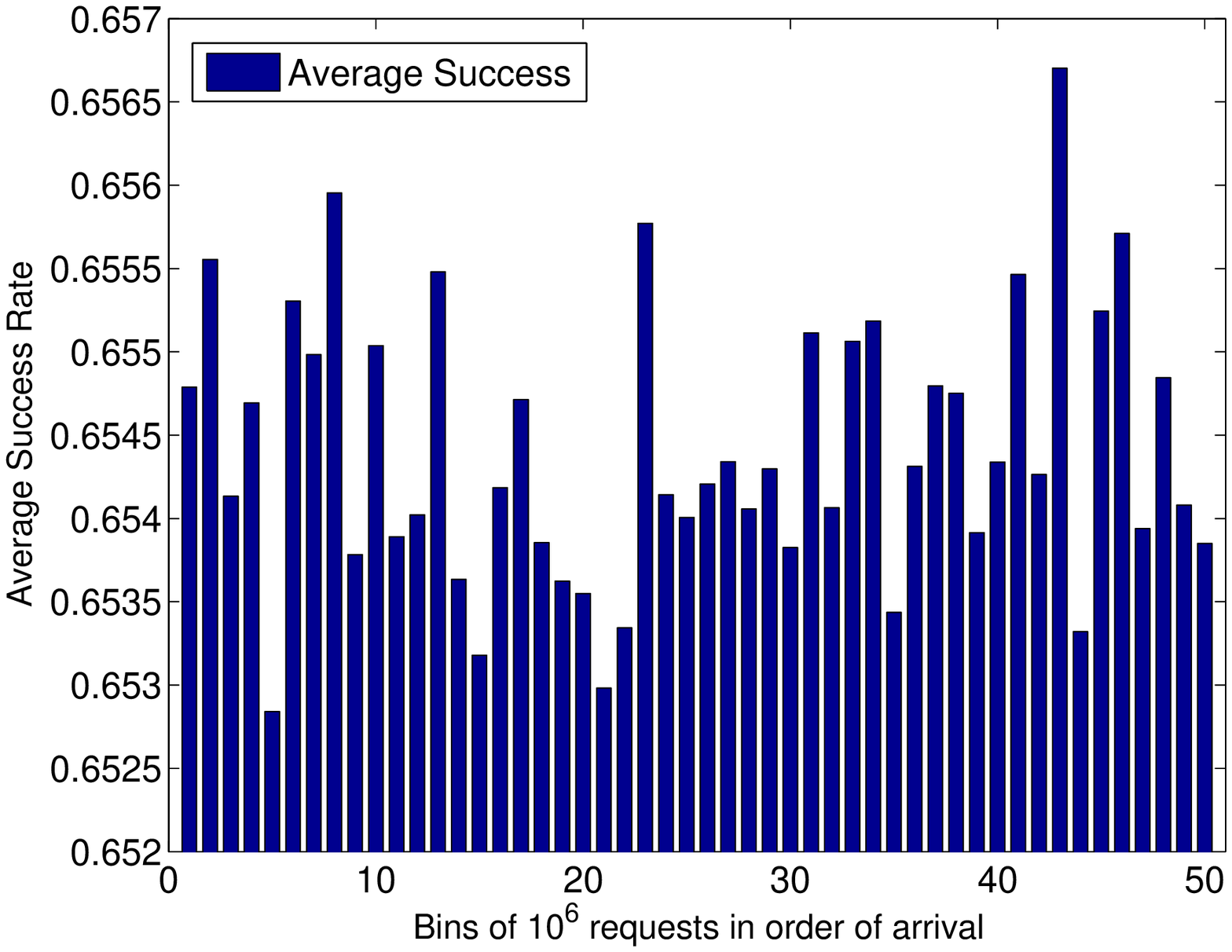}
\includegraphics[width=9.2cm]{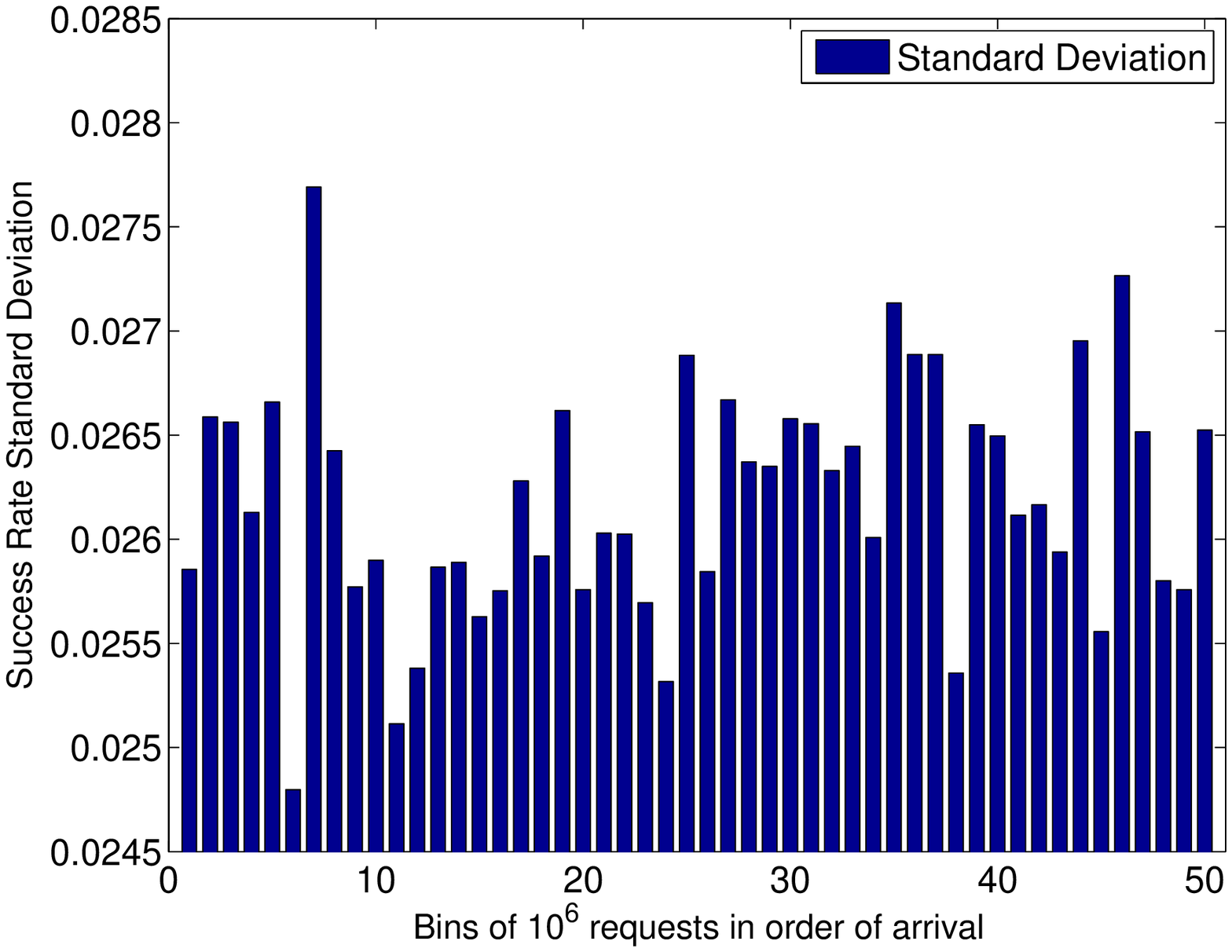}
\end{center}
\caption{Average and standard deviation of the success rate for M1, M2,M3 service requests. Coalitions initiated by a periphery server, {\it Exp4}}
\label{SuccessRatePeriphery}
\end{figure*}

\medskip

The large number of core servers force us to consider several simplifying assumptions: (i) we assume that all core servers have the same capacity, but they are heterogeneous and thus the cost for providing services can be different; (ii) a request is characterized only by the amount of resources needed, while in a realistic scenario other elements such as deadlines, energy consumption, costs, privacy and security would be considered; (iii) in a realistic scenario a bidder would offer several alternatives with different levels of compliance to  user requirements and cost; sophisticated bidding algorithms would choose the optimal bid, while in our simulation a bid consists only of the cost of providing the service and the lowest bid is always selected; (iv) if there is no bid for a request then that request is added to a queue of unsatisfied requests, rather than being used to trigger a renegotiation of the service. This added level of complexity could be addressed in a future iteration of the system.

We experimented with different: number of core servers, number of service requests,  initial system load, demand for resources, and the modes to create a coalition. For the first four experiments we considered the initial organization resulting from the previous experiment with $m=10$, a cloud with $8,388,608$ core servers and $M=1000$ periphery servers subjected to   $50 \times 10^{6}$ service requests. A service request arrives at a randomly chosen periphery server and has an equal probability of being of modes M1, M2, or M3.

\begin{enumerate}
\item
{\it Exp1} - a lightly loaded system with coalitions initiated by a core server.   Initially, the state of each core server is randomly selected such that the fraction of servers in each state is: $20\%$ - sleep state, $40\%$, $15\%$, and $25\%$  running, in M1, M2, and M3 modes, respectively. The initial load of a core server is uniformly distributed in the range $30\% - 80 \% $ of the server's capacity.  The arrival and service processes have exponential distributions with inter-arrival times $\lambda=1.5$ and, respectively,  $\mu=1.2$ units. The workload required is uniformly distributed in the range $[0.1 - 8.0]$ SCUs (Server Compute Units) and only primary contacts are used to assemble a coalition.
\item
{\it Exp2} - similar to {\it} Exp1 but with the workload uniformly distributed in the $[0.1 - 40.0]$ SCUs and secondary contacts are used to assemble a coalition when the primary list is exhausted.
\item
{\it Exp3} - highly loaded system with coalitions initiated by a core server. Initially, the state of each core server is randomly selected; the fraction of servers in M1, M2, and M3 modes is the same, $33\%$. The initial load of the core servers is uniformly distributed in the range $50\% - 80 \% $ of the server capacity. The arrival processes has an exponential distributions with inter-arrival times $\lambda=1.0$ and the service process has a heavy-tail distribution, a Pareto distribution with  $\alpha=2.0$. The workload required is uniformly distributed in the $[0.1 - 40.0]$ SCUs and only primary contacts are used to assemble a coalition.
\item
{\it Exp4} - similar to {\it Exp3} but a coalition is initiated by a periphery server.
\item
{\it Exp5}- heavily loaded, small scale cloud with the number of core servers, $N=100$, and the number of periphery servers, $M=2$, and coalitions are initiated by the periphery servers and subjected to $1,000$ service requests. The initial load of the core servers is uniformly distributed in the range $70\% - 90 \% $ of the server capacity. The arrival and service processes have exponential distributions with inter-arrival times $\lambda=1.5$ and, respectively,  $\mu=1.2$ units. The workload required is uniformly distributed in the $[0.1 - 8.0]$ SCUs and only primary contacts are used to assemble a coalition.
\item
{\it Exp6} - same as {\it Exp5} but the workload required is uniformly distributed in the $[0.1 - 40.0]$ SCUs. Secondary contacts are used when needed.
\end{enumerate}

The results of {\it Exp1} and  {\it Exp2} for the three delivery modes, M1, M2, and M3 are summarized in Table \ref{LowLoadTable} and are illustrated in Figure \ref{LowLoadFig}. We construct bins of $10^{6}$ requests and compute the success rate for each bin; this success rate is computed as the ratio of the number of successful bids to the number of service requests in the bin.  We call this quantity a {\it rate} because the requests are ordered in time thus, the ratio reflects the number of successes in the interval of time correspond to the arrival of $10^{6}$ requests of a certain type. The results in Table \ref{LowLoadTable} show that the workload required expressed as the number of SCUs has little effect on a lightly loaded system.  The first simulation experiment when the core servers use only their primary contacts took slightly less than $10$ hours, while the second one, when the load was higher and the core servers had to use their secondary contacts, took almost $24$ hours.

\begin{figure*}[!ht]
\begin{center}
\includegraphics[width=9.2cm]{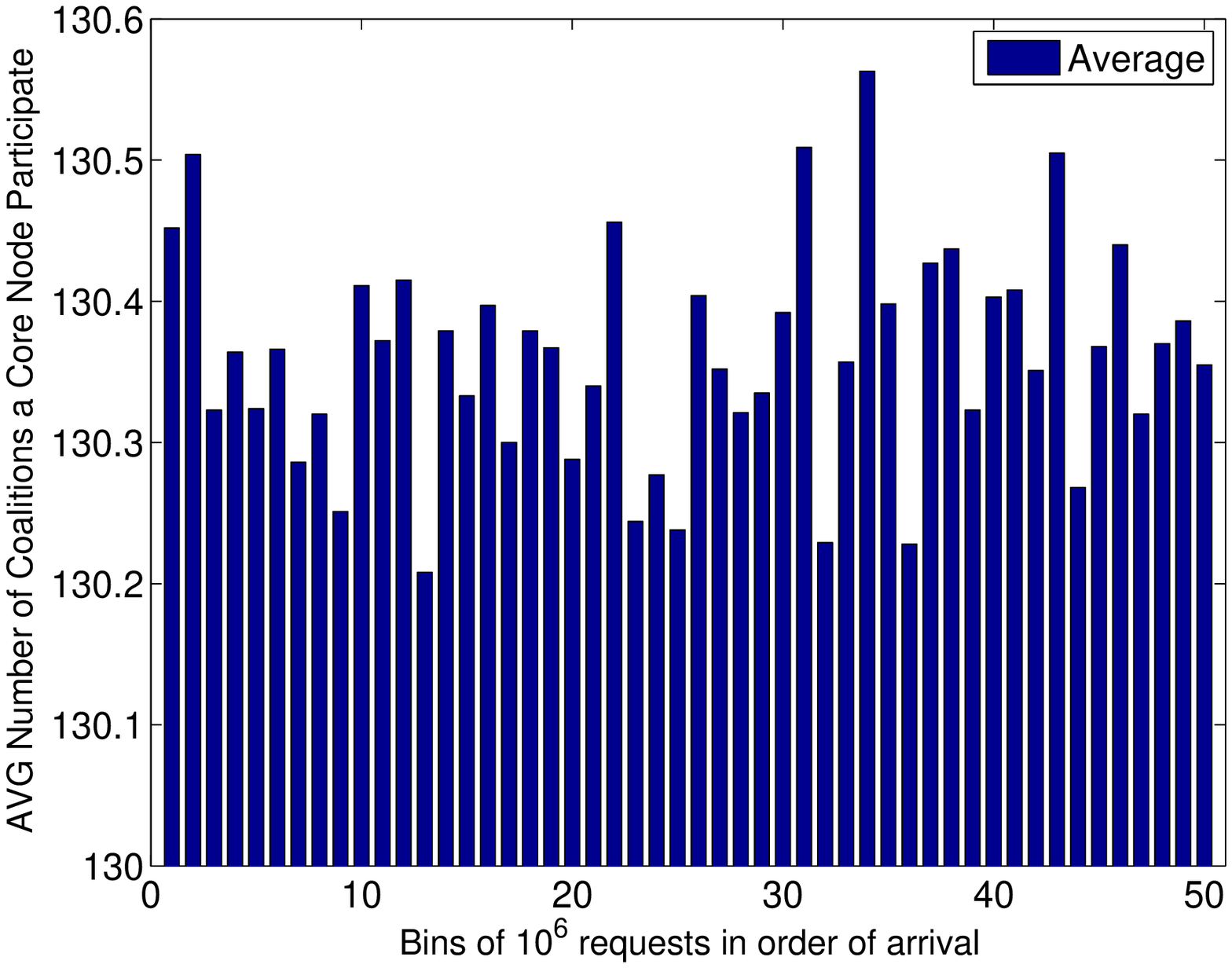}
\includegraphics[width=9.2cm]{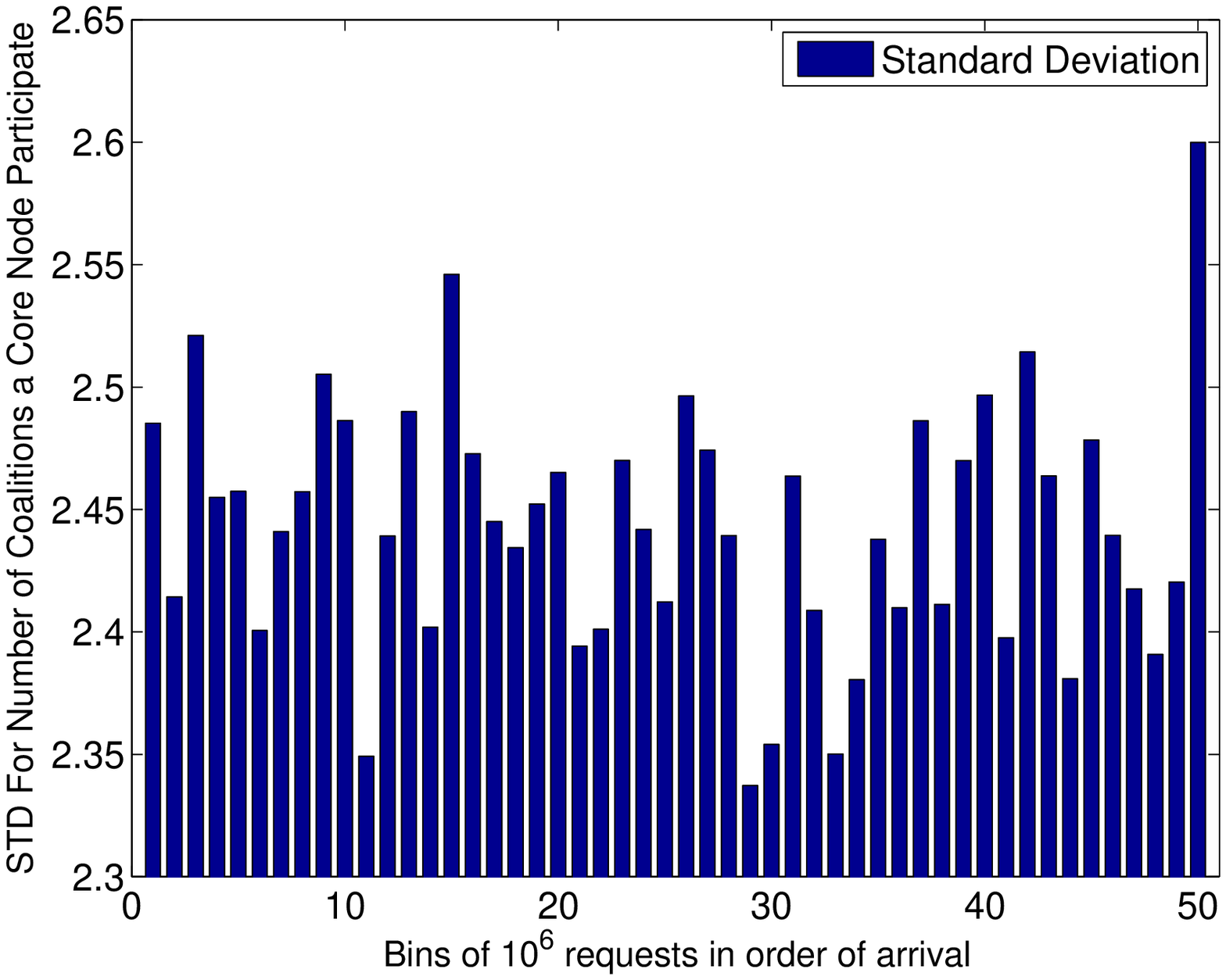}\\
\includegraphics[width=9.2cm]{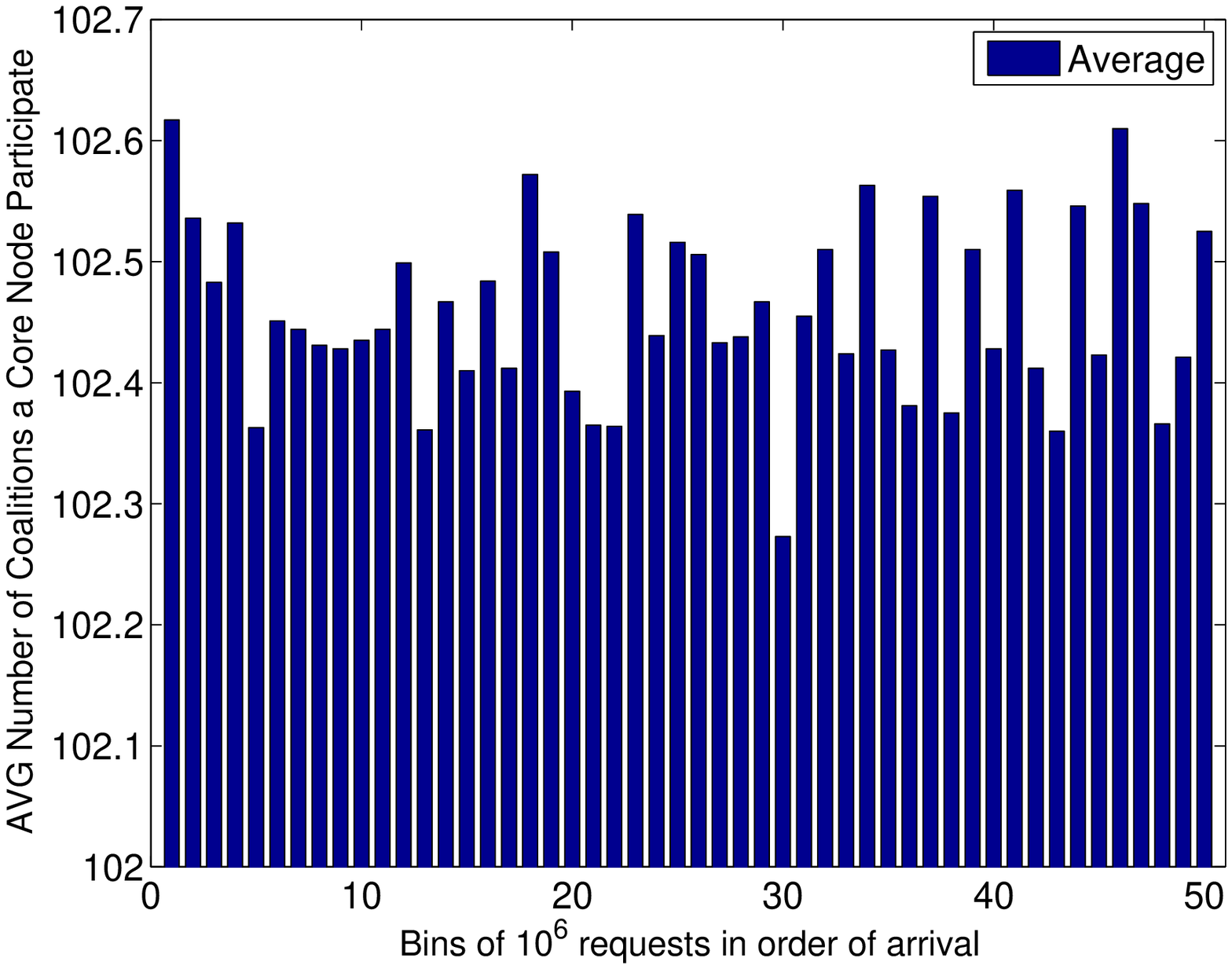}
\includegraphics[width=9.2cm]{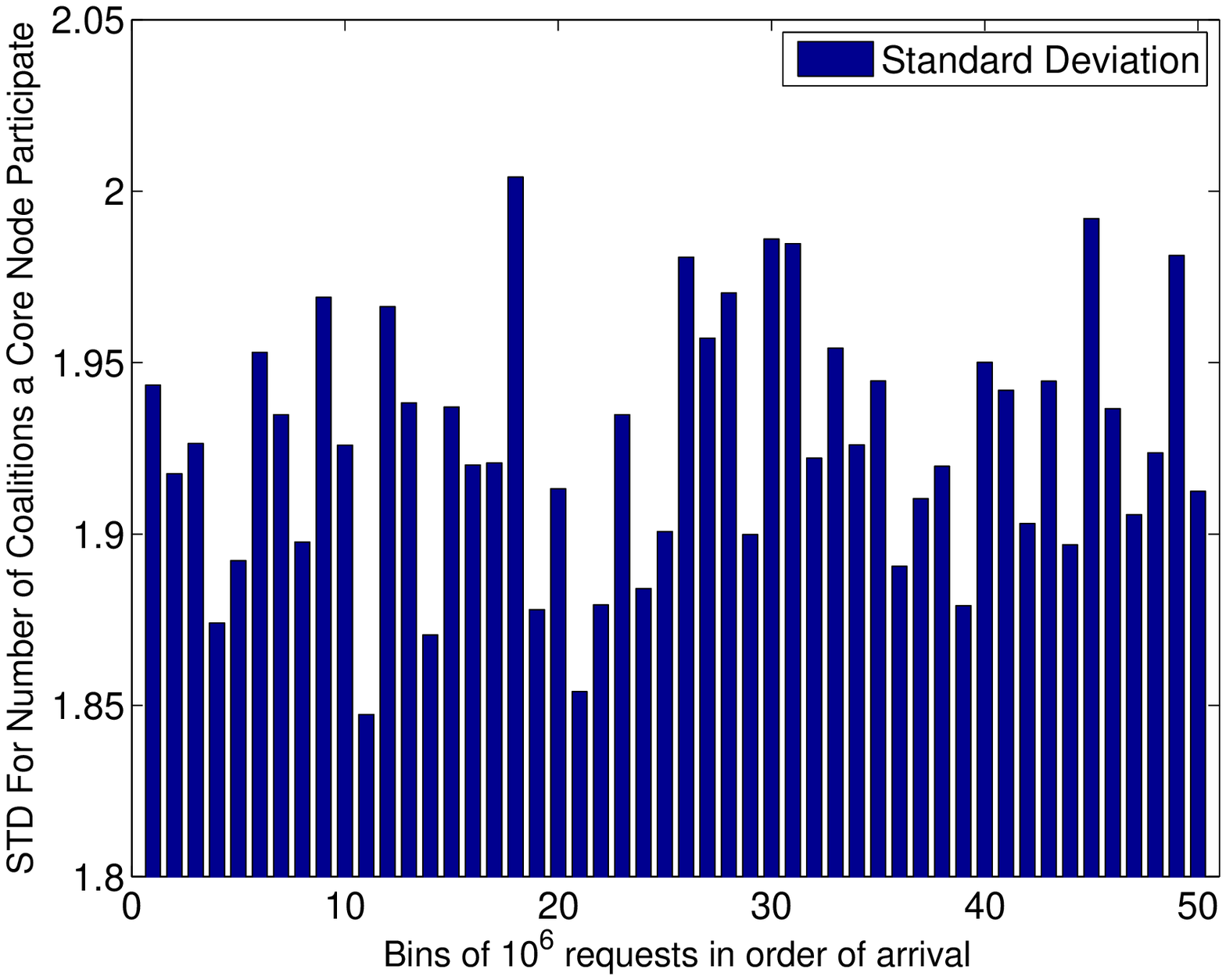}
\end{center}
\caption{Average and standard deviation of number of coalitions a core server is a member of. Coalitions initiated by (top)  a core server. (bottom) a periphery server.}
\label{Coalitions}
\end{figure*}

To illustrate the system behavior for {\it Exp3} and {\it Exp4} we show  histograms of the average success rate and its variance for each of the three modes in Figures \ref{SuccessRateCore} and \ref{SuccessRatePeriphery}, respectively. We split a  $10^{6}$ bin into $10^{3}$ sets with $10^{3}$ requests in each set and compute the standard deviation using the success rate in each set. The success rates range from $65\%$ to close to $80\%$ with relatively small standard deviation for each bin. The core-initiated coalitions have slightly larger success rates because in this mode the leader uses secondary contacts after exhausting the set of primary contacts, therefore has access to a larger population. The average number of coalitions a core server participates in and the standard deviation for these two experiments are shown in Figure \ref{Coalitions}. The averages for the core-initiated mode are slightly larger. The executions times for each one of the two experiments are about $24$ hours.

\begin{figure*}[!ht]
\begin{center}
\includegraphics[width=9.2cm]{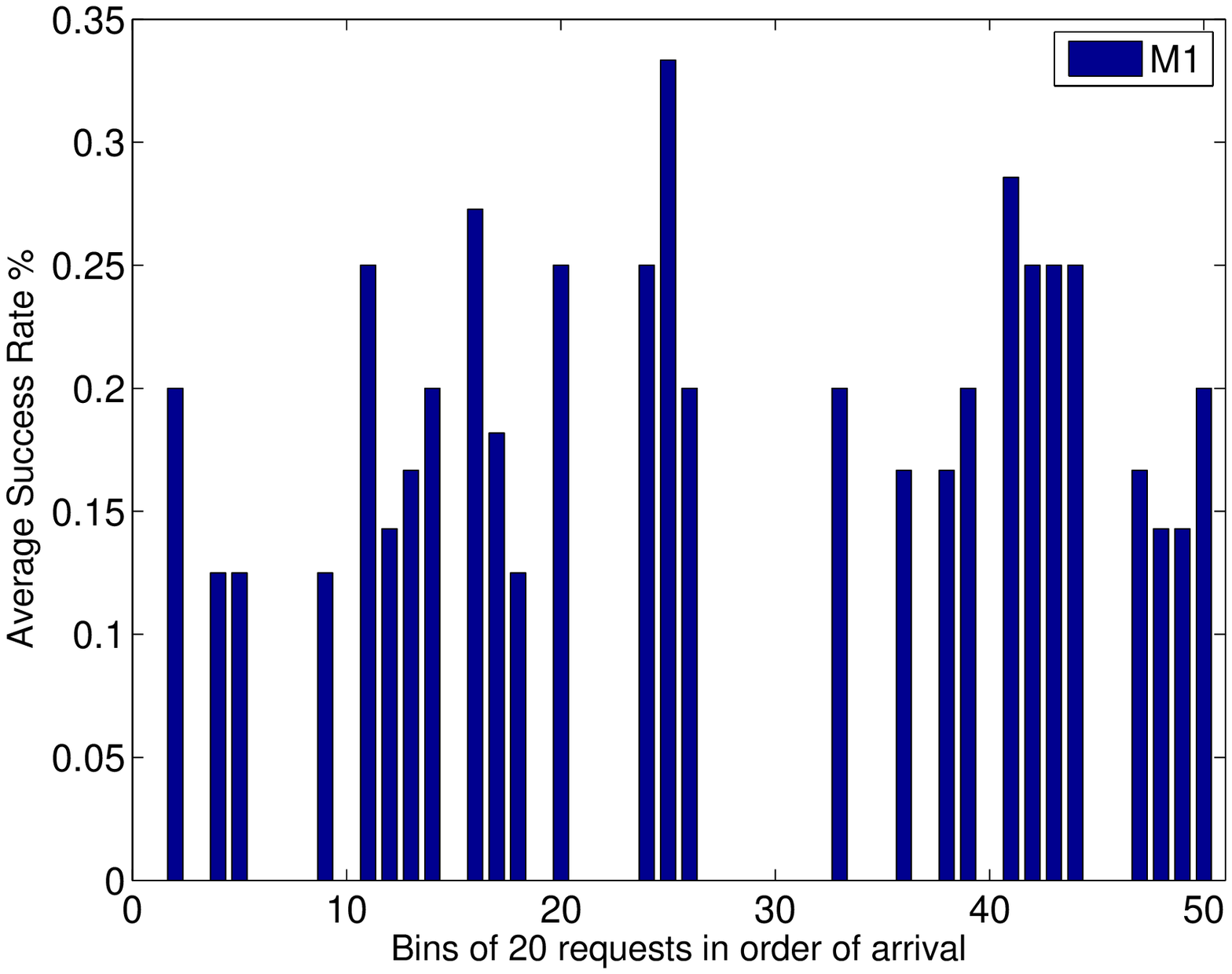}
\includegraphics[width=9.2cm]{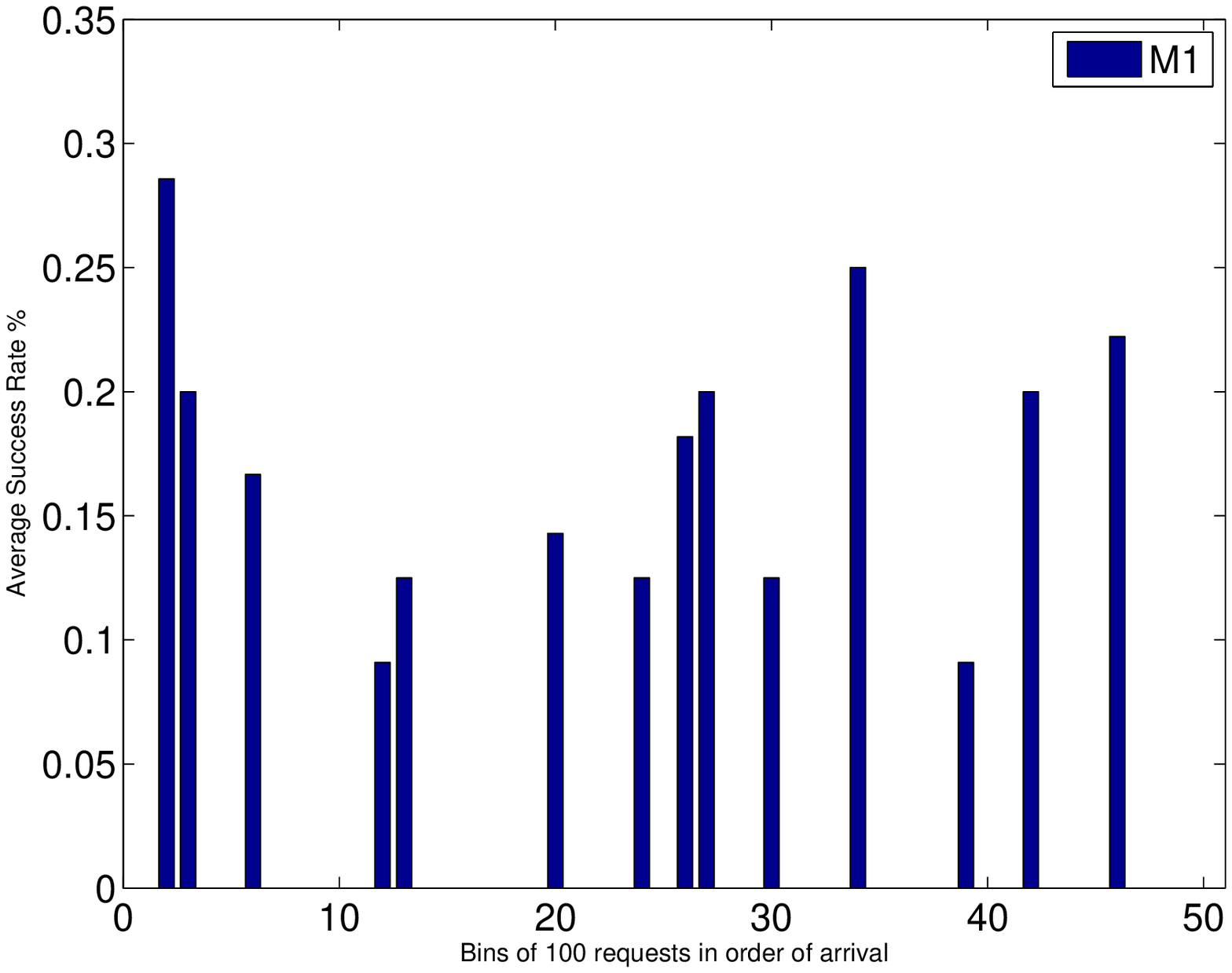}\\
(a)~~~~~~~~~~~~~~~~~~~~~~~~~~~~~~~~~~~~~~~~~~~~~~~~~~~~~~~~~~~~~~~~~~~~~~~(b)~~\\
\includegraphics[width=9.2cm]{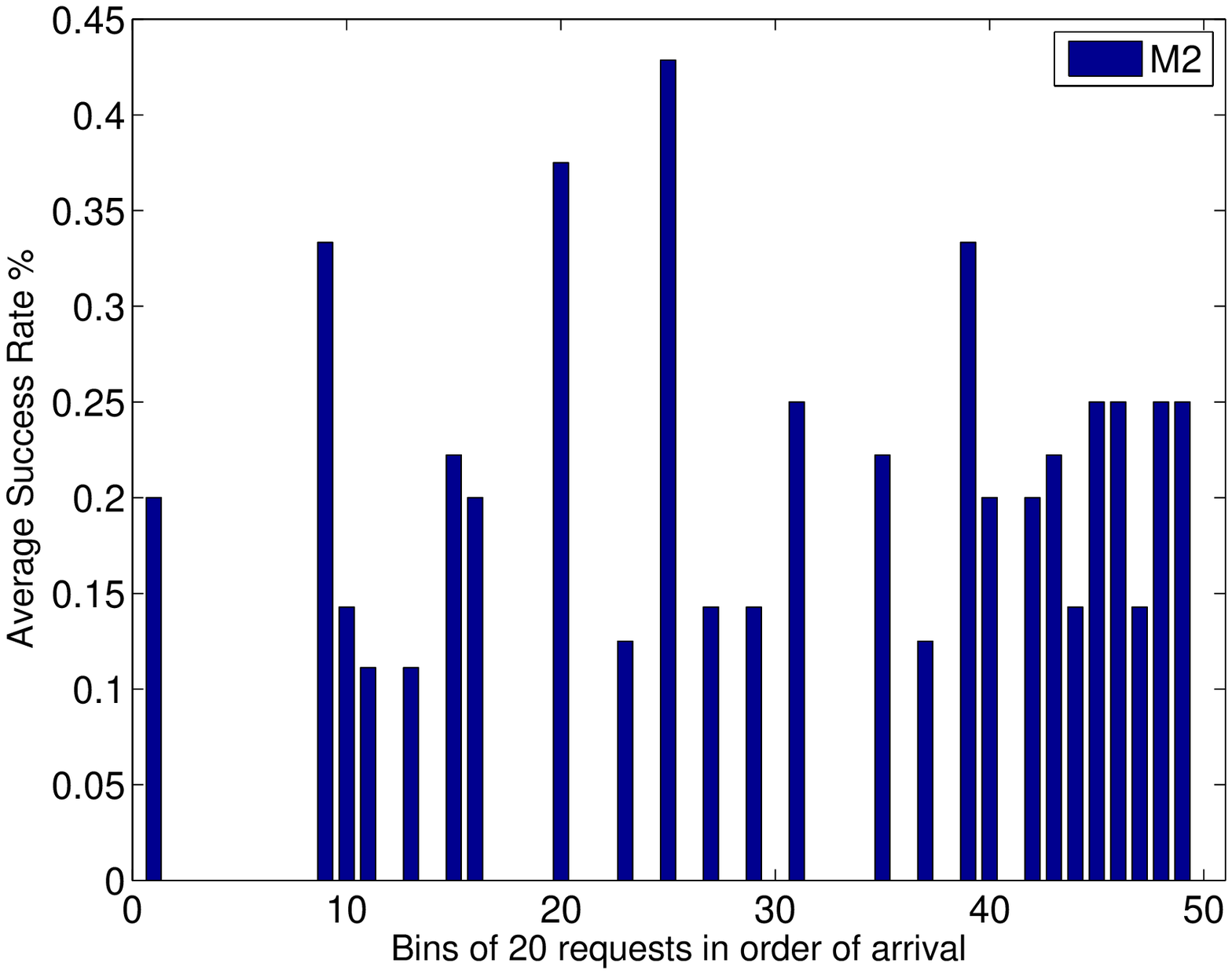}
\includegraphics[width=9.2cm]{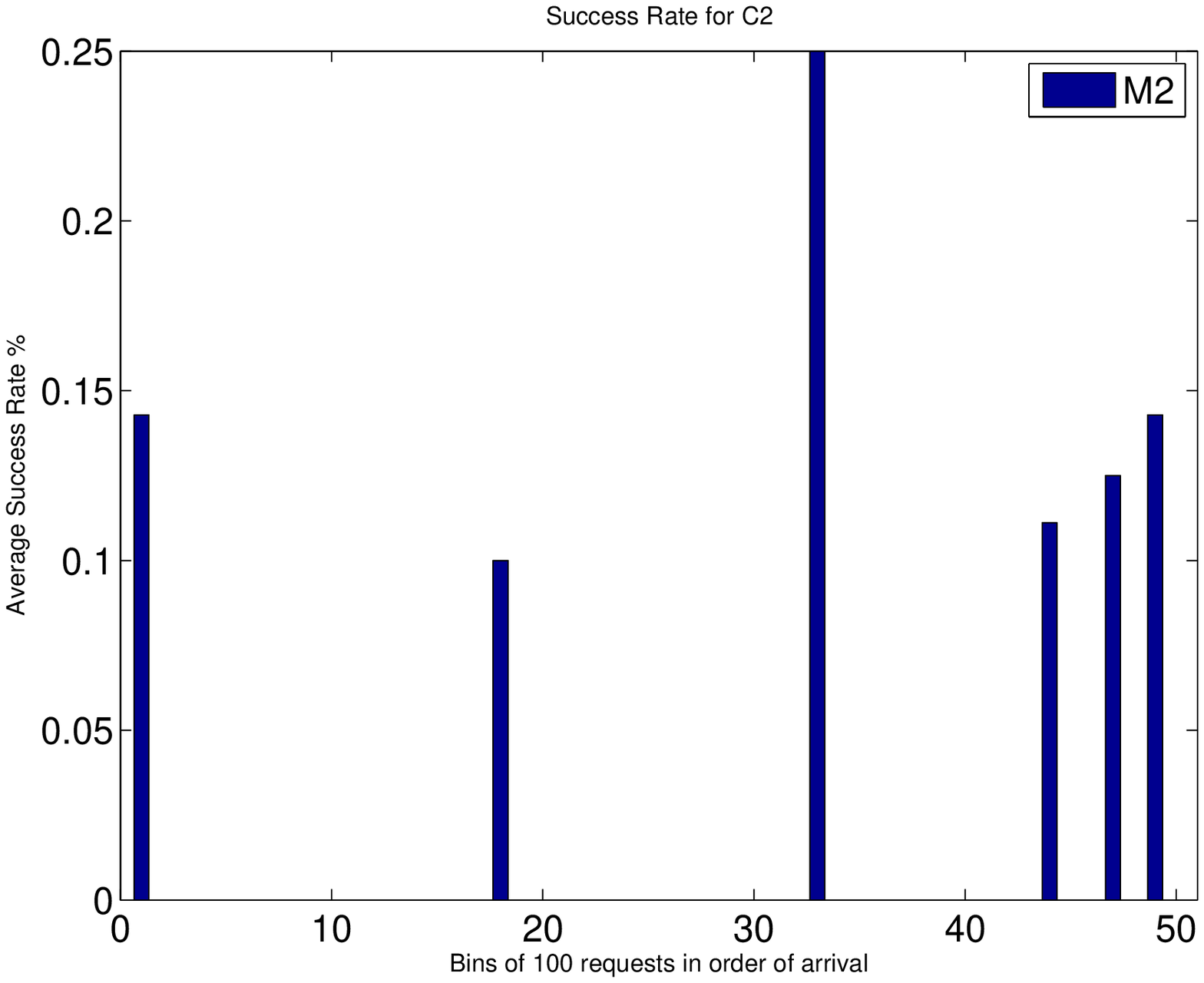}\\
(c)~~~~~~~~~~~~~~~~~~~~~~~~~~~~~~~~~~~~~~~~~~~~~~~~~~~~~~~~~~~~~~~~~~~~~~~(d)~~\\
\includegraphics[width=9.2cm]{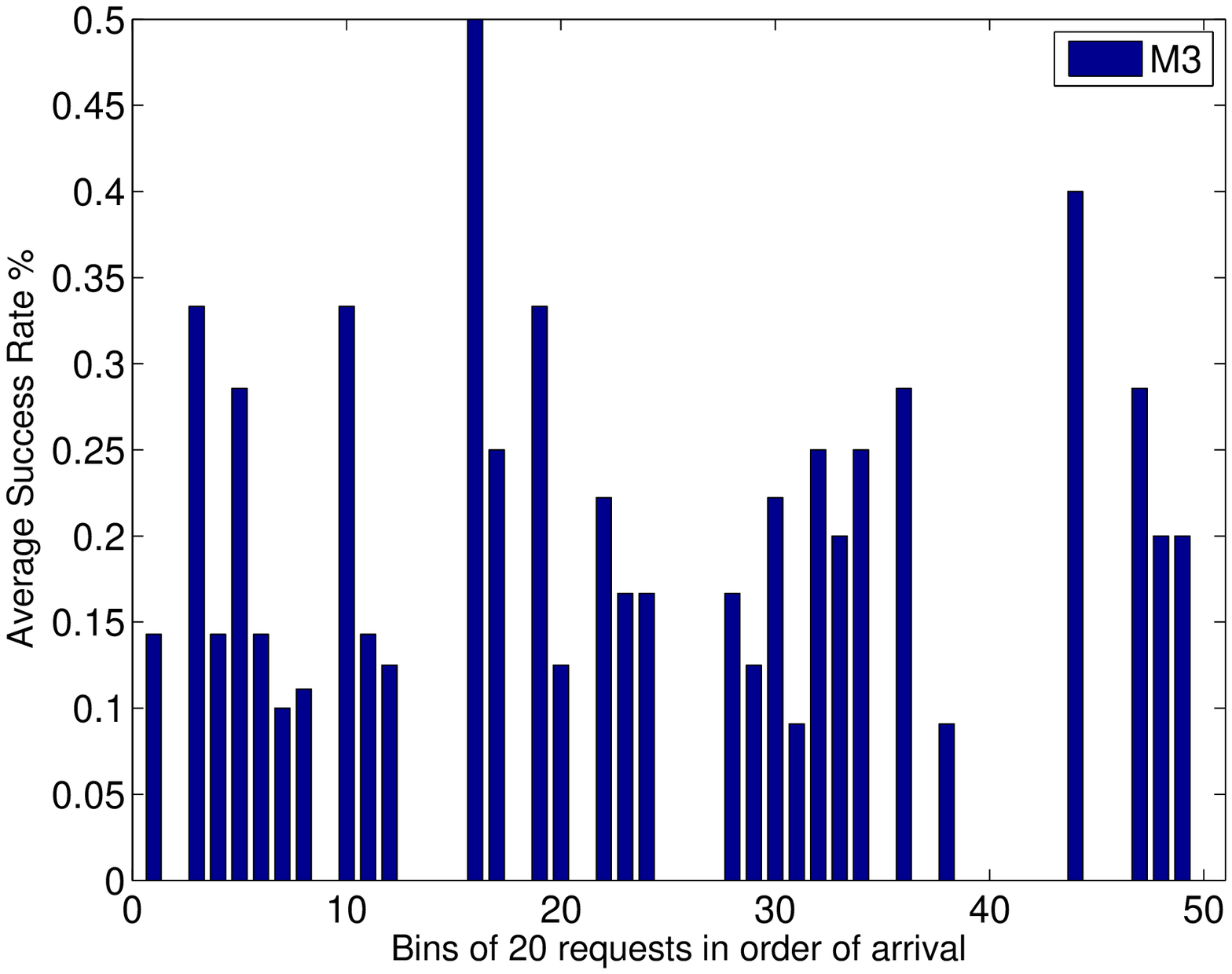}
\includegraphics[width=9.2cm]{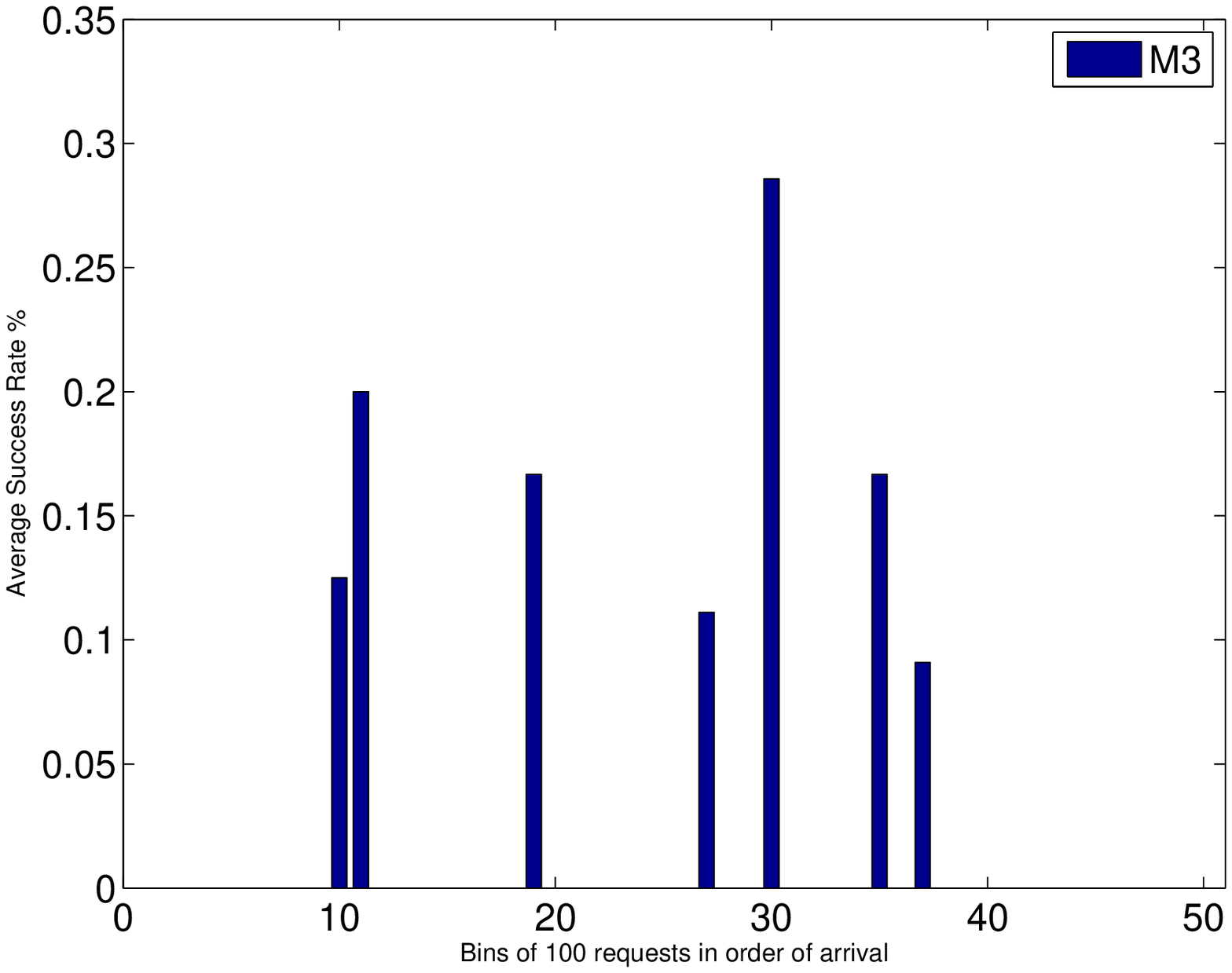}\\
(e)~~~~~~~~~~~~~~~~~~~~~~~~~~~~~~~~~~~~~~~~~~~~~~~~~~~~~~~~~~~~~~~~~~~~~~~(f)
\end{center}
\caption{The ratio of requests and the success rate: (a,b) M1 mode, (c,d) M2 mode, and (e,f) M3 mode. (a,c,e) {\it Exp5}, (b,d,e) {\it Exp6}.}
\label{SmallFig}
\end{figure*}

Lastly  {\it Exp5} and {\it Exp6} show what we expected, namely that in a small, overloaded system, the success rate is low, there are no bids for many requests thus, the admission control works well, Figures \ref{SmallFig}.

\section{Conclusions and Future Work}
\label{Conclusions}
\medskip

The  IaaS cloud delivery model, and the Amazon Web Services in particular, support not only a cost-effective, but also a very convenient and elastic  computing environment. The fact that we are able to simulate the behavior of a complex system with almost ten million components at a cost of slightly more than $\$100$ (the time for the three simulation experiments are $2 + 10 + 24)= 36$ hours and the cost per hour is \$3.0.) gives us an indication of the appeal of cloud computing for scientific and  engineering applications.

Any proposal for a novel cloud computing delivery model has to address the feasibility, the performance, and the cost involved. The first question is whether the state of the art of the technologies involved
allow the development of systems based on the new principles. The next, and the more difficult question, is whether the performance of the proposed systems justifies the investment in the new ideas. This is a more challenging task because qualitative arguments such as security and performance isolation, user-centric organization, self-organization and self-management, support for aggregation, assembly of systems of systems -- clouds of clouds in our context -- have to be balanced against quantitative measures such as revenue, utilization, QoS, and so on.

Numerical simulation is widely used in science and engineering for getting insights into physical phenomena, for checking the accuracy and limitations of theoretical models, for testing hypothesis, or for comparing different design options and parameters of the systems we plan to build. In  all these cases, we start with a model, an abstraction of a physical system or a phenomena, then we carry out the simulation based on this model and, finally, we validate the simulation results by comparing them with theoretical predictions and, whenever feasible, with measurements of the physical system.

How to validate the simulation results for a new model of a complex system, a heterogeneous system with a very large number of components and with many interaction channels among them? In this case we have only the model, there is no physical system, and no comprehensive theory describing the system as a whole. One can only validate the algorithms which control the evolution of the system; we can do that by showing that the model of the system subject to different stimuli reacts according to the behavior prescribed by the algorithm. Admittedly, this is a weak form of validation of simulation results, but the only one available. Of course, one could test the algorithms on a small-scale system, but scalability is an insidious problem, and there is no guarantee that an idea working beautifully on a small-scale system will work at all at a larger scale.

For this reason in this paper we limit our discussion to qualitative rather than quantitative results and  two aspects of the feasibility of the self-organizing model: the initial system organization and the bidding mechanisms. The model is rather complex and an in-depth analysis in necessary to determine optimal parameters such as: the ratio between the core and periphery servers, number of primary and secondary contacts, the size of the population invited to place bids in response to a service requests.  We use several simplifying assumptions discussed in Section \ref{Simulation}, as well as, parameters we believe reflect a ``typical'' behavior for the simulation experiments. While, during the initial system configuration, we used a random selection process, more intricate learning algorithms are likely to lead to a more effective selection of primary contacts by each core server.

The simulation experiments we conducted show that the initial system organization phase can be tuned to provide each core server with a balanced number of primary and secondary contacts. The bidding mechanisms seem to work well when the workload is relatively low. In spite of the larger overhead, core-initiated coalition formation seems more effective than periphery-initiated coalitions.

As our objective is a qualitative, rather than a quantitative analysis we do not compute confidence intervals for our simulation results. Our primary goal was to test the scalability of the self-organizing architecture and the simulation experiments for a system with a close to nine million core servers took almost 36 hours on a very powerful cloud configuration. The cost for a large number of runs necessary to report confidence intervals was prohibitive.

The disruptive approach we propose poses a number of new questions to be addressed by future research. For example, some of the open research questions related to the role of single system image techniques in self-organizing clouds are: (i) Can the distributed hypervisors be modified to support dynamic horizontal scaling? (ii) What are the performance differences between distributed hypervisors and best-of-breed SSI operating systems? (iii) How do SSI techniques, such as process and thread migration, perform when used over with low-latency networking interconnect technologies such as RDMA Verbs \cite{Hilland03}? (iv) Can abstract cloud service descriptions be efficiently mapped to static heterogeneous clusters of multicore, GPU, FPGA and MIC processing resources? (v) Can dynamic horizontal scaling of coprocessor resources be integrated into abstract cloud service descriptions?

Further work includes the development of a cloud service description language and of efficient bidding algorithms which take into account the workload, deadlines, energy consumption, costs, and possibly other parameters of a service request. An interesting concept developed by Papadimitriou and his collaborators is that of ``mixability'' the ability of an entity to interact with others \cite{Livnat08}. The authors argue that mixability is
a critical element in the evolution of species, a good indicator of the fitness and of the ability of an individual to transmit the positive traits to its descendants. In our system mixability will quantify the ability of a server to successfully interact with other servers in each of the three modes of operation.

Lastly, some of the highly desirable features of the architecture we introduce, such as security and privacy due to isolation of the servers assigned to a specific application, come at a higher cost for providing the service. The main appeal of the solution we propose is the potential to satisfy a very broad range of user preferences and allow service level agreements that reflect the specific user requirements and the contractual obligation of the CSP for a specific service and user. This becomes  increasingly more important as information privacy laws differ from country to country and a CSP has to obey the security and privacy rules and laws demanded by each user.

\section{Literature}
\renewcommand{\refname}{}


\begin{thebibliography}{11}

\bibitem{Abbott07}
R.~Abbott.
\newblock ``Complex systems engineering: putting complex systems to work.''
\newblock {\it Complexity,} {\bf 13}(2):10--11, 2007.

\bibitem{Ardagna07}
D.~Ardagna, M.~Trubian, and L.~Zhang.
\newblock ``SLA based resource allocation policies in autonomic environments.''
\newblock {\it J. Parallel Distrib. Comp.}, {\bf 67}(3):259--270, 2007.

\bibitem{Ardagna11}
D.~Ardagna, B.~Panicucci, M.~Trubian, and L.~Zhang.
\newblock ``Energy-aware autonomic resource allocation in multi-tier virtualized environments.''
\newblock {\it IEEE Trans. on Services Computing,} {\bf 5}(1):2--19, 2012.


\bibitem{Arrasjid13}
J.  Y. Arrasjid et. al.
\newblock {\it VMware vCloud Architecture Toolkit.}
\newblock VMware Press, Upple Saddle River NJ, 2013.


\bibitem{Auty10}
M.~Auty, S.~Creese, M.~Goldsmith, and P.~Hopkins.
\newblock ``Inadequacies of current risk controls for the cloud.''
\newblock {\it Proc. IEEE 2nd Int. Conf. on Cloud Computing Technology and Science,}
pp. 659--666, 2010.

\bibitem{Azambuja10}
M.~Azambuja, R.~Pereira, K.~Breitman, and M.~Endler.
\newblock ``An architecture for public and open submission systems in the cloud.''
\newblock {\it Proc. IEEE 3rd Int. Conf. on Cloud Computing}, pp. 513--517, 2010.

\bibitem{Balduzzi12}
M. Balduzzi, J. Zaddach, D. Balzarotti, E. Kirda, and S. Loureiro.
\newblock ``A security analysis of Amazon's elastic compute cloud service.''
\newblock {\it Proc. 27th Annual ACM Symp. on Applied Computing}, pp. 1427--1434, 2012.

\bibitem{Barak11}
A. Barak  and A. Shiloh.
\newblock ``The {Virtual OpenCL (VCL)} cluster platform.''
\newblock {\em Proc. Intel European Research \& Innovation Conference}, Leixlip, Ireland, pp. 196--200, 2011.


\bibitem{Barak12}
A. Barak and A. Shiloh.
\newblock {\em The {MOSIX} Cluster Operating System for High-Performance
  Computing on {Linux} Clusters, Multi-Clusters and Clouds}, 2012.

\bibitem{Barham03}
P. Barham, B. Dragovic, K. Fraser, S. Hand, T. Harris, A. Ho, R. Neugebauer, I. Pratt, and A. Warfield.
\newblock ``Xen and the art of virtualization.''
\newblock {\it Proc. SOSP'03, 19th ACM Symp. on Operating Systems Principles}, pp. 164--177, 2013.

\bibitem{Bernstein11}
D.~Bernstein, D.~Vij, and S. Diamond.
\newblock ``An Intercloud cloud computing economy - technology, governance, and market blueprints.''
\newblock {\it Proc. SRII Global Conference,} pp. 293--299, 2011.

\bibitem{Brandic10}
I.~Brandic, S.~Dustdar, T.~Ansett, D.~Schumm, F.~Leymann, and R.~Konrad.
\newblock ``Compliant cloud computing (C3): Architecture and language support for user-driven compliance management in clouds.''
\newblock {\it Proc. IEEE 3rd Int. Conf. on Cloud Computing}, pp. 244--251, 2010.

\bibitem{Buyya97}
R. Buyya.
\newblock ``Single system image: need, approaches, and supporting {HPC} systems.''
\newblock In {\em Proc. Int. Conf. on Parallel and Distributed
  Processing Techniques and Applications (PDPTA'97)}.
Las Vegas, Nevada, USA, pp. 1106--1113, 1997.

\bibitem{Buyya01}
R. Buyya, T. Cortes, and H. Jin.
\newblock ``Single system image.''
\newblock {\em International Journal of High Performance Computing
  Applications\/}~{\em 15,\/}~2, 124--135, 2001.


\bibitem{Celesti10}
A. Celesti, F. Tusa, M. Villari, and A.Puliafito.
\newblock ``How to enhance cloud architecture to enable cross-federation.''
\newblock {\it Proc. IEEE 3rd Int. Conf. on Cloud Computing}, pp. 337--345, 2010.


\bibitem{Chandra03}
A.~Chandra, P.~Goyal, and P.~Shenoy.
\newblock ``Quantifying the benefits of resource multiplexing in on-demand data centers.''
\newblock {\it Proc. 1-st Workshop on Algorithms and Architecture for Self-Managing Systems,} 2003.

\bibitem{Chang10}
V.~Chang, G.~Wills, and D.~De Roure.
\newblock ``A review of cloud business models and sustainability.''
\newblock {\it Proc. IEEE 3rd Int. Conf. on Cloud Computing}, pp. 43--50, 2010.


\bibitem{Chapman05}
M. Chapman and G. Heiser.
\newblock ``Implementing transparent shared memory on clusters using virtual machines.''
\newblock In {\em Proc. USENIX Annual Technical Conference}, pp. 23--23,  2005.

\bibitem{Cofer08}
H. Cofer, M. Fouquet-Lapar, T. Gamerdinger, C. Lindahl,
  B. Losure, A, Mayer, J. Swoboda,  and T. Utsumi.
\newblock ``Creating the world's largest reconfigurable supercomputing system
  based on the scalable {SGI}{\textregistered} {Altix}{\textregistered} 4700
  system infrastructure and benchmarking life-science applications.''
\newblock {\em Reconfigurable Computing: Architectures, Tools and
  Applications\/}, pp. 268--273, 2008.

\bibitem{Colp11}
P. Colp, M. Nanavati, J. Zhu, W. Aiello, G. Coker, T. Deegan, P. Loscocco, and A. Warfield.
\newblock ``Breaking up is hard to do: security and functionality in a
commodity hypervisor.''
\newblock {\it Proc. Symp. Operating Systems Principles,} pp. 189--202, 2011.

\bibitem{Cramton06}
P.~Cramton, Y.~Shoham, and R.~Steinberg, Eds.,
\newblock {\it Combinatorial Auctions,} MIT Press, 2006.

\bibitem{CSA11}
Cloud Security Alliance.
\newblock ``Security guidance for critical areas of focus in cloud computing V3.0.''
\newblock {\it https://cloudsecurityalliance.org/guidance}, 2011.

\bibitem{Dutreilh10}
X.~Dutreild, N.~Rivierre, A.~Moreau, J.~Malenfant, and I.~Truck.
\newblock ``From data center resource allocation to control theory and back.''
\newblock {\it Proc. IEEE 3rd Int. Conf. on Cloud Computing}, pp. 410--417, 2010.

\bibitem{Gandhi11}
A. Gandhi, and M. Harchol-Balter.
\newblock ``How data center size impacts the effectiveness of dynamic power management.''
\newblock {\it Proc. 49th Annual Allerton Conference on Communication, Control, and Computing, Urbana-Champaign}, pp. 1864--1869, 2011.

\bibitem{Gandhi12}
A. Gandhi, M. Harchol-Balter, R. Raghunathan, and M.Kozuch.
\newblock ``AutoScale: dynamic, robust capacity management for multi-tier data centers.''
\newblock {\it ACM Trans. on Computer Systems}, {\bf 30}(4):1--26,  2012.

\bibitem{Gandhi12a}
A. Gandhi, T. Zhu, M. Harchol-Balter, and M.Kozuch.
\newblock ``SOFTScale: stealing opportunistically for transient scaling.''
\newblock {\it Proc. Midlleware 2012}, in Lecture Notes in Computer Science, Springer Verlag, Berlin, Heidelberg, pp. 142--163, 2012.

\bibitem{Gandhi12b}
A. Gandhi, M. Harchol-Balter, R. Raghunathan, and M.Kozuch.
\newblock ``Are sleep states effective in data centers?''
\newblock {\it Proc. Int. Conf. on Green Computing}, pp. 1--10, 2012.


\bibitem{Garfinkel05}
S.~Garfinkel and M. Rosenblum.
\newblock ``When virtual is harder than real: security challenges in virtual machines based computing environments.''
\newblock {\it Proc. Conf. Hot Topics in Operating Systems,} pp. 20--25, 2005.

\bibitem{Gmach09}
D.~Gmach, J.~Rolia, and L.~Cerkasova.
\newblock ``Satisfying service-level objectives in a self-managed resource pool.''
\newblock {\it Proc. 3rd. Int. Conf. on Self-Adaptive and Self-Organizing Systems,}
pp. 243--253, 2009.

\bibitem{Gruschka10}
N.~Gruschka and M.~Jensen.
\newblock ``Attack surfaces: A taxonomy for attacks on cloud services.''
\newblock {\it Proc. IEEE 3rd Int. Conf. on Cloud Computing}, pp. 276--279, 2010.

\bibitem{Gupta09}
V.~Gupta and M.~Harchol-Balter.
\newblock ``Self-adaptive admission control policies for resource-sharing systems.''
\newblock {\it Proc. 11th Int. Joint Conf. Measurement and Modeling Computer Systems (SIGMETRICS'09)}, pp. 311--322, 2009.


\bibitem{Gell-Mann88}
M. Gell-Mann.
\newblock ``Simplicity and complexity in the description of nature.''
\newblock {\it Engineering and Sciences}, LI, vol. 3, California Institute of Technology,
pp. 3-9, 1988.

\bibitem{Hilland03}
J. Hilland, P. Culley, J. Pinkerton and R. Recio,
\newblock ``RDMA Protocol Verbs Specification.''
\newblock {\it RDMAC Consortium Draft Specification},
2003

\bibitem{Hinchey06}
M. Hinchey, R. Sterritt, C. Rouff, J. Rash, W. Truszkowski.
\newblock ``Swarm-based space exploration.''
\newblock {\it  ERCIM News} 64, 2006.

\bibitem{Hopfield82}
J. Hopfield.
\newblock ``Neural networks and physical systems with emergent collective computational abilities.''
\newblock {\it Proc. National Academy of Science,}  79, pp. 2554-2558, 1982.

\bibitem{Kaneda05}
K. Kaneda, Y. Oyama, and A. Yonezawa.
\newblock ''A virtual machine monitor for providing a single system image.''
\newblock {\em Proc. 17th IPSJ Computer System Symposium
  (ComSys’ 05)}.
pp. 3--12, 2005.

\bibitem{Intel13}
Intel.
\newblock {\it Intel{\textregistered} Xeon Phi{\texttrademark} Coprocessor System Software Developers Guide},
2013

\bibitem{Kalyvianaki09}
E.~Kalyvianaki, T.~Charalambous, and S.~Hand.
\newblock ``Self-adaptive and self-configured CPU resource provisionong for virtualized servers using Kalman filters.''
\newblock {\it Proc. 6th Int. Conf. Autonomic Comp. (ICAC2009),} pp. 117--126, 2009.

\bibitem{Kephart07}
J.~Kephart, H.~Chan, R.~Das, D.~Levine, G.~Tesauro, F.~Rawson, and C.~Lefurgy.
\newblock ``Coordinating multiple autonomic managers to achieve specified power-performance tradeoffs.''
\newblock {\it Proc. 4th Int. Conf. Autonomic Computing (ICAC2007),} pp. 100-109, 2007.


\bibitem{Krugman96}
P.R. Krugman.
\newblock ``The Self-organizing Economy.''
\newblock {\it Blackwell Publishers,} 1996.

\bibitem{Kusic08}
D.~Kusic, J.~O.~Kephart, N.~Kandasamy, and G.~Jiang.
\newblock ``Power and performance management of virtualized computing environments via lookahead control.''
\newblock {\it Proc. 5th Int. Conf. Autonomic Comp. (ICAC2008)}, pp. 3--12, 2008.



\bibitem{Lamport98}
L.~Lamport.
\newblock ``The part-time parliament.''
\newblock{\it ACM Trans. on Computer Systems} {\bf 2}:133--169, 1998.

\bibitem{Lamport01}
L.~Lamport.
\newblock ``Paxos made simple.''
\newblock{\it ACM SIGACT News} {\bf 32}(4):51--58, 2001.


\bibitem{Lim09}
H C.~Lim,  S.~Babu, J. S.~Chase, and S. S.~Parekh.
\newblock ``Automated control in cloud computing: challenges and opportunities.''
\newblock {\it Proc. First Workshop on Automated Control for Datacenters and Clouds,},
ACM Press, pp. 13--18, 2009.

\bibitem{Lin12}
M. Lin, Z. Liu, A. Wierman, and L. L. H. Andrew
\newblock ``Online algorithms for geographical load balancing.''
\newblock {\it Proc. Int. Conf. on Green Computing}, pp. 1--10, 2012.


\bibitem{Livnat08}
A. Livnat, C. Papadimitriou, J. Dushoff, and M. W. Feldman.
\newblock ``A mixability theory of the role of sex in evolution.''
\newblock {\it Proc. National Academy of Science,} {\bf 105}(50):19803-19807, 2008.


\bibitem{Lu11}
Y.-Y. Lu, J.-J. Slotino, and A. L. Barab\'asi.
\newblock ``Controllability of complex networks.''
\newblock {\it Nature} 473, 167, 2011.

\bibitem{Li10}
C.~Li, A.~Raghunathan, and N.~K.Jha.
\newblock ``Secure virtual machine execution under an untrusted management OS.''
\newblock {\it Proc. IEEE 3rd Int. Conf. on Cloud Computing}, pp. 172--179, 2010.

\bibitem{MAAS}
\newblock ``Metal as a service.''
\newblock {\it https://maas.ubuntu.com/}, accessed on October 19, 2013.


\bibitem{Margery03}
D. Margery, G. Vall{\'e}e, R. Lottiaux, C. Morin,
  and J.-Y. Berthou.
\newblock ``{Kerrighed}: A {SSI} cluster {OS} running {OpenMP}.''
\newblock Research Report RR-4947, INRIA, 2003.

\bibitem{Marinescu08}
D. C. Marinescu, X. Bai, L.~B{\"o}l{\"o}ni, H. J.
Siegel, R. E. Daley, and I-J. Wang.
\newblock ``A macroeconomic model for resource allocation in large-scale distributed systems.''
\newblock {\it Journal of Parallel and Distributed Computing}, vol. 68, pp. 182--199, 2008.

\bibitem{Marinescu09}
D. C. Marinescu, J. P. Morisson , and H. J. Siegel.
\newblock ``Options and Commodity Markets for Computing Resources.''
\newblock  {\it Market Oriented Grid and Utility Computing,} R. Buyya and K. Bubendorf, Eds., Wiley,
ISBN: 9780470287682, pp. 89--120, 2009.


\bibitem{Marinescu10}
D. C.~Marinescu,  C.~Yu, and G. M.~Marinescu.
\newblock {``Scale-free, self-organizing very large sensor networks.''} Journal of Parallel and Distributed Computing (JPDC), {\bf 50}(5):612--622, 2010.

\bibitem{Marinescu13}
D. C. Marinescu.
\newblock ``Cloud Computing; Theory and Practice.''
\newblock {\it Morgan Kaufmann}, 2013.

\bibitem{Menon05}
A.~Menon, J. R.~Santos, Y.~Turner, G. J.~Janakiraman, and W.~Zwaenepoel.
\newblock ``Diagnosing performance overheads in Xen virtual machine environments.''
\newblock {\it Proc. First ACM/USENIX Conf. on Virtual Execution Environments,}  2005.

\bibitem{Morin04}
C. Morin, P. Gallard, R. Lottiaux and G. Vall{\'e}e.
\newblock ''Towards an efficient single system image cluster operating system.''
\newblock {\em Future Generation Computer Systems\/}~{\em 20,\/}~4, pp.505--521, 2004.

\bibitem{Noel13}
P-A. No\"el, C. D. Brummitt, and R. M. D\'~Souza.
\newblock {``Controlling self-organizing dynamics on networks using models that self-organize.''} \newblock {\it Phys. Rev. Lett.} {\bf 111}, 078701, 2013.

\bibitem{Osinski09}
P. Osi{\'n}ski and E. Niewiadomska-Szynkiewicz.
\newblock ``Comparative study of single system image clusters.''
\newblock {\em Evolutionary Computation and Global Optimization}, pp. 145--154, 2009.


\bibitem{Paya13}
A. Paya and D. C. Marinescu.
\newblock ``Energy-aware application scaling on a cloud.''
\newblock http://arxiv.org/pdf/1307.3306v1.pdf, July 2013.

\bibitem{Pearson10}
S. Pearson and A. Benameur.
\newblock ``Privacy, securitry, and trust  issues arising from cloud computing.''
\newblock {\it Proc. Cloud Computing and Science,} pp. 693--702, 2010.

\bibitem{Peng08}
J. Peng, X. Long, and L. Xiao.
\newblock ``{DVMM}: A distributed {VMM} for supporting single system image on
  clusters.''
\newblock{\em Proc. 9th
  International Conference forYoung Computer Scientists},
pp. 183 --188, 2008.

\bibitem{Pfister98}
G. Pfister.
\newblock {\em In Search of Clusters: The Ongoing Battle in Lowly Parallel
  Computing}.
\newblock Prentice-Hall, Inc, 1998.

\bibitem{Pfister09}
G. Pfister.
\newblock ``Multi-multicore {Single System Image}/{Cloud Computing}. {A} good
  idea?''
\newblock {\em The Perils of Parallel (Blog)}.
\newblock 2009. Online; accessed September 26 2012.

\bibitem{Price08}
M. Price.
\newblock ``The paradox of security in virtual environments.''
\newblock {\it Computer,} {\bf 41}(11):22--28, 2008.

\bibitem{Rosenblum05}
M.~Rosenblum and T.~Garfinkel.
\newblock ``Virtual machine monitors: Current technology and future trends.''
\newblock {\it Computer,} {\bf 38}(5):39--47, 2005.

\bibitem{Sarathy10}
V. Sarathy, P. Narayan, and R. Mikkilineni.
\newblock ``Next generation cloud computing architecture.''
\newblock {\it Proc. IEEE Int. Workshops on Enabling Technologies: Infrastructures for Collaborative Enterprises}, pp. 48--53, 2010.

\bibitem{Schuster07}
P. Schuster.
\newblock ``Nonlinear dynamics from Physics to Biology. Self-organization: An
old paradigm revisited.''
\newblock {\it Complexity,} {\bf 12}(4):9-11, 2007.

\bibitem{Sommerville12}
I. Sommerville, D. Cliff, R. Calinescu, J, Keen, T. Kelly, M. Kwiatowska, J. McDermid, and R. Paige.
\newblock ``Large-scale IT complex systems.''
\newblock {\it Communications of the ACM}, {\bf 55}(7):71--77, 2012.

\bibitem{Sugerman01}
J.~Sugerman, G.~Venkitachalam, and B.~Lim.
\newblock ``Virtualizing I/O devices on VMware Workstation's hosted virtual machine monitor.''
\newblock {\it Proc. USENIX Conf.}, pp. 70--85, 2001.

\bibitem{Stokely10}
M.~Stokely, J.~Winget, E.~Keyes, C.~Grimes, and B.~Yolken.
\newblock ``Using a market economy to provision compute resources across planet-wide clusters.''
\newblock {\it Proc. Int. Parallel and Distributed Processing Symp. (IPDPS 2009)}, pp. 1--8, 2009.


\bibitem{Tung07}
C.~Tung, M.~Steinder, M.~Spreitzer, and G.~Pacifici.
\newblock ``A scalable application placement controller for enterprise data centers.''
\newblock {\it Proc. 16th Int. Conf. World Wide Web (WWW2007),} 2007.

\bibitem{Turing52}
A.M. Turing.
\newblock ``The chemical basis of morphogenesis.''
\newblock {\it Philosophical Transactions of the Royal Society of London,} Series B 237:37-72, 1952.

\bibitem{Van10}
H.~N.~Van, F.~D.~Tran, and J.-M.~Menaud.
\newblock ``Performance and power management for cloud infrastructures.''
\newblock {\it Proc. IEEE 3rd Int. Conf. on Cloud Computing}, pp. 329--336, 2010.


\bibitem{Walker99}
B. Walker and D. Steel.
\newblock ``Implementing a full single system image {UnixWare} cluster:
  Middleware vs. Underware.''
\newblock {\em Proc. Intl Conf. on Parallel and Distributed
  Processing Techniques and Applications (PDPTA)}, 1999.

\bibitem{Walker03}
B. Walker.
\newblock {\it Open single system image ({openSSI}) {Linux} cluster project}.
\newblock Technical report, 2003.

\bibitem{Wang09}
X. Wang, M. Zhu, L. Xiao, Z. Liu, X. Zhang, and X. Li.
\newblock ``{NEX}: Virtual machine monitor level single system support in {Xen}.''
\newblock {\em First Int. Workshop on Education Technology and Computer Science (ETCS'09)},
Vol.~3. pp. 1047--1051, 2009.

\end{thebibliography}
\end{document}